\documentclass[%
preprint, 
nofootinbib,
 amsmath,amssymb,
 aps, physrev,
]{revtex4-2}

\usepackage{graphicx}% Include figure files
\usepackage{dcolumn}% Align table columns on decimal point
\usepackage{bm}% bold math
\newcommand{\eqn}[1]{\begin{equation} #1 \end{equation}}

\newcommand{\deriv}[2]{\ensuremath{\frac{d#1}{d#2}}}

\usepackage{mathptmx}
\usepackage{appendix}
\usepackage{etoolbox}
\usepackage{graphicx}
\usepackage{amsmath}
\usepackage{url}
\usepackage{algorithmic}
\usepackage{float}%If you want a figure to be in a particular spot use: \begin{figure}[H]
\usepackage{wasysym}%For diameter symbol
\usepackage{threeparttable}%For allowing footnotes in tables using the threeparttable environment
\usepackage{overpic}%To write stuff on top of pictures
\usepackage{psfrag}%To replace text in the EPS files with LaTeX code
\usepackage[normalem]{ulem}
\usepackage{tikz}
\usepackage{placeins}
\usepackage{setspace}
\usepackage[hang,flushmargin]{footmisc}

\begin{document}

\preprint{APS/123-QED}

\title{\textbf{Emergent Workload Inequality in Collective Excavation} }% 

\author{Laura K. Treers}
\affiliation{Department of Mechanical Engineering, University of Vermont, Burlington, VT, USA}
 
\author{Aradhya Rajanala}%
\affiliation{Interdisciplinary Graduate Program in Quantitative Biosciences, Georgia Institute of Technology, Atlanta, GA USA}%

\author{Nathan Nguyen}
\affiliation{%
Division of Physics, Mathematics and Astronomy (PMA), California Institute of Technology, Pasadena, CA USA}%

\author{Naomi Wagner}
\affiliation{%
School of Physics, Oglethorpe University, Atlanta, GA USA}%

\author{Michael A. D. Goodisman}
\affiliation{%
School of Biological Sciences, Georgia Institute of Technology, Atlanta, GA USA}%

\author{Daniel I. Goldman}
\affiliation{%
School of Physics, Georgia Institute of Technology, Atlanta, GA USA}%

\date{\today}% It is always \today, today,

\begin{abstract}

Living collectives and artificial swarms frequently employ a division of labor, wherein individuals take on different tasks or perform different amounts of work. However, the mechanisms used by collectives to divide labor remain poorly understood. Here, we study how workload inequality arises in collectives by monitoring excavation in \textit{Solenopsis invicta} fire ants, whose coordination in constrained environments makes them an attractive system for studying division of labor. We vary group size (between 2 and 25 ants) and track digging activity to create Lorenz curves and corresponding Gini coefficients, which represent relative workload inequality. We find that that workload becomes more unequal as group size increases: the number of ``active" ants scales with the square root of the group size.
We implement a cellular automata (CA) model in which agents regulate their activity based on local crowding in the tunnel. The CA reproduces experimental Gini coefficients over a wide range of parameters and group sizes, indicating that local decisions emergently account for the scaling of workload inequality. 
An analytic rate equation model recovers the square root scaling with the assumption that individuals exit the tunnel at a rate which scales quadratically with the group size. 
Power law scalings in workload distribution have been observed in other systems, including social and natural sciences; however, these laws are primarily observational. Here, we provide a mechanistic explanation for the emergent workload scaling patterns in constrained biological collectives, offering insight into organization in both natural and future task capable engineered collectives and swarms.

\end{abstract}

\maketitle

%\tableofcontents

\section{Introduction} 
Biological and robotic collective systems succeed because many individual entities work together to complete complex tasks. Individuals in some biological collectives operate in a decentralized capacity, using local information to inform decision making. Nevertheless, simple rules followed by many individuals often lead to complex, effective group behaviors; for example, flocks of birds \cite{reynolds1987flocks} and schools of fish \cite{couzin2003self} coordinate their movements to conserve energy, evade predation, and share information \cite{moussaid2009collective}. The lens of active matter physics has been used to describe these biotic systems across scales, from molecular and cellular self organization \cite{needlemanactivematter} to organismal swarms \cite{deblais2023worm}. Specifically, active matter physics can describe macroscopic phenomena as an emergent result of individual behavior, through the analysis of environmental interactions and constraints \cite{RamaswamyActiveMatter, paoluzzi2024flocking, goldmanactivematter, vernerey2019biological, jung2025kinetic, cocconi2025dissipation}. Such constraints include space restrictions, in the case of collective construction \cite{aguilar_collective_2018}, or limited resources, in the case of group foraging \cite{detrain_self-organized_2006}.  
Throughout evolutionary history, biological systems have evolved strategies for maintaining effective collective behaviors despite the presence of these constraints.

% Collective systems often operate under the presence of constraints: such constraints include space restrictions, in the case of collective construction \cite{aguilar_collective_2018}, or limited resources, in the case of group foraging \cite{detrain_self-organized_2006}. Throughout evolutionary history, biological systems have evolved strategies for maintaining effective collective behaviors despite the presence of these constraints.

One general principle employed by collective systems operating under constraints is division of labor \cite{gordon2016division, robinson2009division, jeanson2019within}. Division of labor occurs when individuals within a society specialize and have distinct roles or undertake different amounts of work.  Observed variation in individual contributions in work effort could arise through a variety of factors across collective systems \cite{holbrook2011division, charbonneau_lazy_2015, toth2005worker, huang1992honeybee}. However, the mechanisms leading to specialization in individual behavior and division of labor among individuals remain poorly understood \cite{west_major_2015}. Even in human social sciences, high inequality in labor has been observed and modeled with various scaling laws, such as Price's law \cite{price1963little}, which was originally used to describe the distribution of productive work in academic publishing. These scaling observations for labor distribution in human social systems currently lack mechanistic explanation. 

Division of labor has been shown to be an important factor in the enormous success of social insects, such as ants, termites, some bees, wasps, thrips, beetles, and aphids \cite{oster_caste_1978, beshers_models_2001}.
% Division of labor is quite common AND HYPOTHESIZED RESPONSIBLE FOR THE ENORMOUS SUCCESS [XX CAN WE SAY SUCH A THING, I'M SURE THERE IS A REVIEW] OF  social insects like ants, termites, some bees, wasps, thrips, beetles, and aphids. 
These social insects represent only two percent of insect taxa but account for more than half of total insect biomass \cite{wilson_success_1990, bourke_social_1995}. In such insect collectives, individuals coordinate activities and ultimately behave as a single ‘superorganism’ \cite{queller_kin_1998, wilson_social_1971, strassmann_insect_2007}.
% One of the main reasons social species are so successful is the high level of division of labor within colonies \cite{charbonneau_when_2015, beshers_models_2001, smith_workload_2022}. 
One family of social insects, Formicidae, or ants, represent one of many independent examples of the evolution of sociality in insects. Similar to many other social insects, ants live in large colonies and must coordinate to achieve various tasks necessary for survival, such as nest construction. 

\begin{figure}
    \centering
    \includegraphics[width=0.7\linewidth]{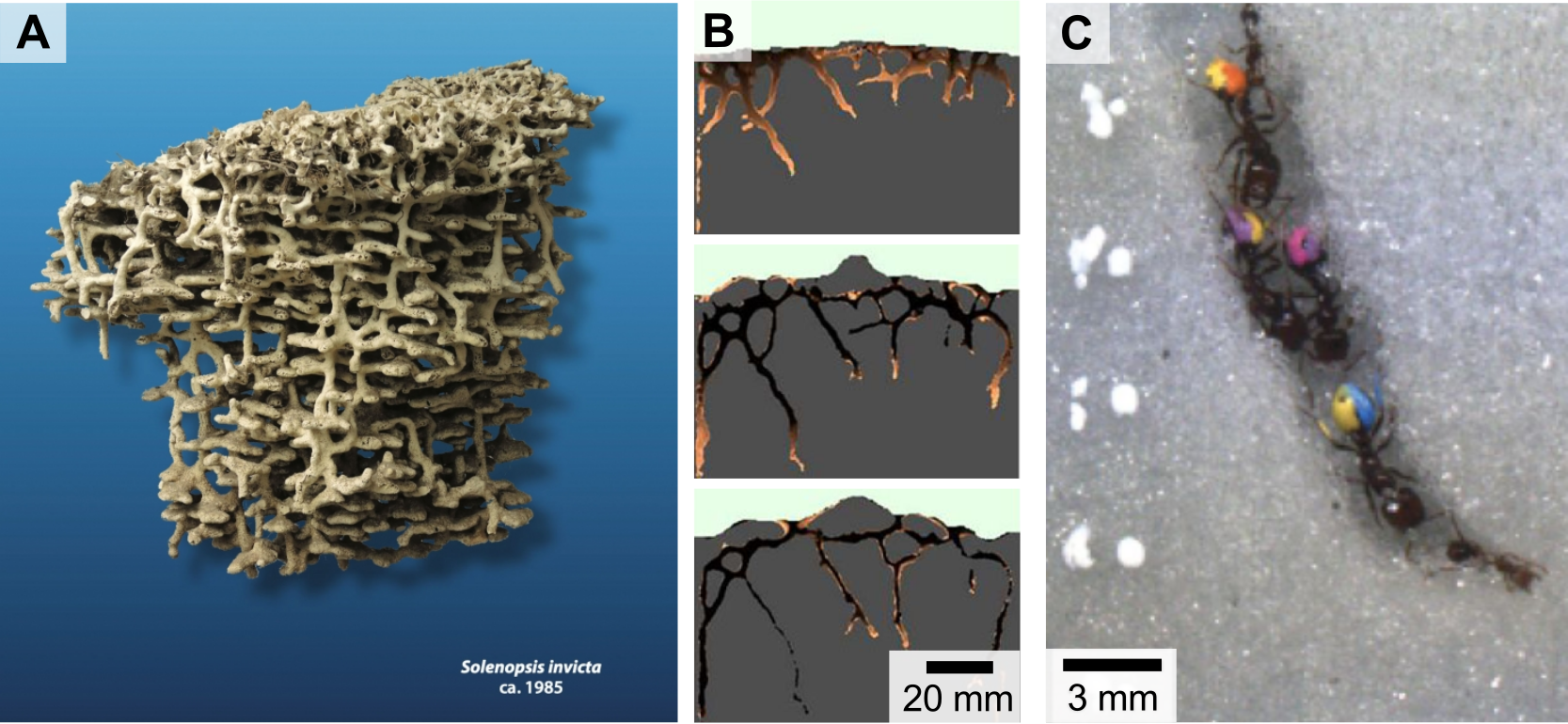}
    \caption{\textbf{Image of fire ant nest construction}. \textbf{(A)} Cast of a fire ant (\textit{Solenopsis invicta}) nest, created by Walter Tschinkel. The nest is approximately 40 cm in diameter, and consists of a complex array of narrow, branching tunnels (adapted from \cite{tschinkel2021ant}).  \textbf{(B)} Tunnel excavation in a 2D ``ant farm'' experiment, over 9 hours, adapted from \cite{avinery_agitated_2023}. Each image represents a 3 hour slice of ant activity. Tunnel color represents first-exploration time, ranging from darkest for earliest and brightest for latest activity within the time slice. \textbf{(C)} A group of fire ants excavates grains at the end of a narrow tunnel (adapted from \cite{aguilar_collective_2018}). 
    }
    \label{fig:nest_intro}
\end{figure}

Nests are fundamentally important to the success of ant species, and the behaviors leading to the successful construction of these nests are remarkable \cite{invernizzi_deconstructing_2019, gordon_ecology_2019, wenzel_14_1991, theraulaz_origin_1998, hansell_animal_2005, theraulaz_formation_2003, camazine_self-organization_2020, perna_when_2017, moffett_ant_2021}: 
thousands of low-level individual interactions result in functional, space filling structures \cite{theraulaz_origin_1998, detrain_self-organized_2006, sumpter_principles_2005, fewell_social_2003, couzin_collective_2009, bardunias_queue_2010, invernizzi_deconstructing_2019, khuong_stigmergic_2016}.
Fire ants (\textit{Solenopsis invicta}) excavate nests comprised of a network of subsurface branching tunnels (Figure \ref{fig:nest_intro}A) in diverse  substrates worldwide \cite{monaenkova_behavioral_2015, gravish_effects_2012, gravish_climbing_2013, gravish_glass-like_2015, holldobler_ants_1990}. This nest construction necessitates division of labor: individuals must coordinate activities to successfully manipulate and shape the environment \cite{scheiner_editorial_2024, duarte_evolutionary_2011, oster_caste_1978, beshers_regulation_2024}. Furthermore, fire ant nest construction is readily observable in the lab, allowing for accessible observations of workload distribution. Thus, fire ants are an attractive system for studying division of labor.

Prior studies demonstrated that individual contributions to the single task of nest excavation are highly unequal, both in laboratory and in field studies \cite{charbonneau_lazy_2015}. In fact, a large percentage of a colony of workers at any given time does not participate in excavation \cite{charbonneau_workers_2015, charbonneau_when_2015}. In fire ant excavation, 30\% of workers complete approximately 70\% of the total work \cite{aguilar_collective_2018}. Recent studies posited that this behavior is adaptive: using both cellular automata as well as robophysical models, Aguilar et al. showed that fire ants' unequal workload distribution optimized traffic flow, reducing jams and increasing excavation speed \cite{aguilar_collective_2018}. 

Despite the ubiquity of workload inequality, how this inequality arises from local interactions during nest construction is poorly understood. 
Some studies have argued that certain ants are less intrinsically motivated to work, such that they ``specialize'' in inactivity \cite{charbonneau_workers_2015, charbonneau_who_2017}. Others have argued that workers are not intrinsically different; rather, other factors, such as environmental conditions, create a self-regulating phenomenon in which some ants choose to remain inactive. These self-regulating phenomena have recently been probed via mathematical models \cite{feng_dynamics_2021, khajehnejad_explaining_2023, porfiri_socially_2024, napoli2025role}. These studies have largely remained in simulated environments and focused on colony-level behavioral trends. Additionally, many of these studies do not center the role of environmental constraints, which may heavily influence collective dynamics \cite{aguilar_collective_2018}.

We hypothesize that the mechanisms underlying workload inequality might be driven by local interactions in the tight tunnels in which the ants operate (Figure \ref{fig:nest_intro}B-C). These tunnels are consistently narrow (approximately 2 ant body widths wide) across a range of substrates; constructing narrow tunnels has been hypothesized as a means of offloading locomotor control to the environment \cite{gravish_climbing_2013}. 
In this paper, we study small groups of fire ants in such constrained environments to understand the mechanisms underlying workload inequality. 
We observe that inactivity scales sublinearly with group size; in particular, we show that the number of active ants in a given group scales as the square root of the total group size. Using both experimental and cellular automata simulation methods, we argue that self-regulation of activity may be modulated by encounter rate or relative density of individuals in a tunnel. Through an analytic model, we then show how a square root scaling in workload can emerge in constrained collective systems. 

\section{Experimental Design}

We hypothesize that by studying behavior of small (10$^{1}$  - 10$^{2}$) groups of individuals in the crowded environment of early nest construction, we can study the pressures that lead to inactivity in these smaller groups. We can then study how small group activity results in widespread workload inequality in the larger collective. 
Prior studies which utilized a small but fixed number of individuals \cite{aguilar_collective_2018} could not probe the mechanisms underlying this inequality. 
Instead, we vary the group size in a fixed experimental area to systematically understand the role of crowding and confinement in workload inequality \cite{fewell2016scaling}. Specifically, we study different sized groups of \textit{Solenopsis invicta} fire ants, known for coordination in highly constrained environments \cite{tschinkel_fire_2013}. 

% XX SET UP YOUR HYPOTHESIS AND RATIONALE FOR THE EXPERIMENTS HERE (PULL FROM ABOVE PARAGRAPH IN INTRO)

\begin{figure}[]
    \centering
    \includegraphics[width=.65\linewidth]{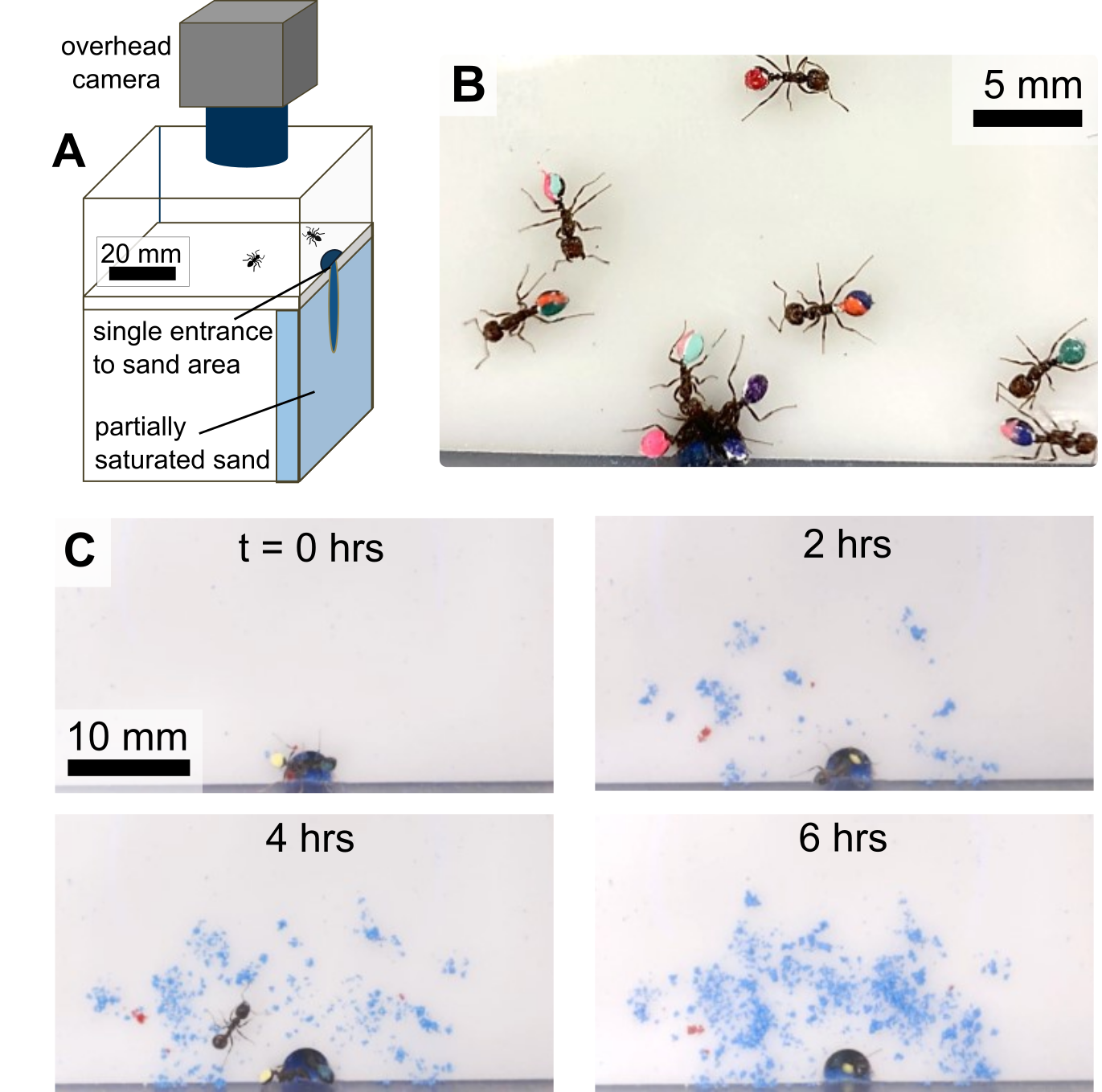}
    \caption{\textbf{Painting ants enables individual ant tracking}. \textbf{(A)} Diagram of camera and container used to record simultaneous experimental trials.  \textbf{(B)} Painted and color-coded ants used in trials with more than 10 individuals. \textbf{(C)} Camera view of trial at 4 different points in time over 6 hours. 
    }
    \label{fig:exp setup 1}
\end{figure}

Fire ant colonies were collected from locations in the metro Atlanta, GA area. The colonies were maintained in a climate controlled room on the Georgia Tech campus. Colonies were fed a protein-carbohydrate mixture 2x weekly and supplied a constant water source via cotton-covered tubes. To study digging behavior, plastic containers ($\approx$65 x 65 mm), shown in Figure \ref{fig:exp setup 1}A, were prepared such that a small region of width $\approx$2 mm along one edge was filled with partially saturated colored sand (saturated 20\% by mass, grain size distribution 75 - 400 $\mu m$). A 4 mm diameter hole on the side of a plastic covering enabled ants to enter the sand but restricted movement in and out of this region to a single location. This setup enabled monitoring of when each ant entered and exited the artificial ``nest'' region, and thus these observations served as a proxy for each ant's activity. 

\begin{figure*}[b]
    \centering
    \includegraphics[scale=0.75]{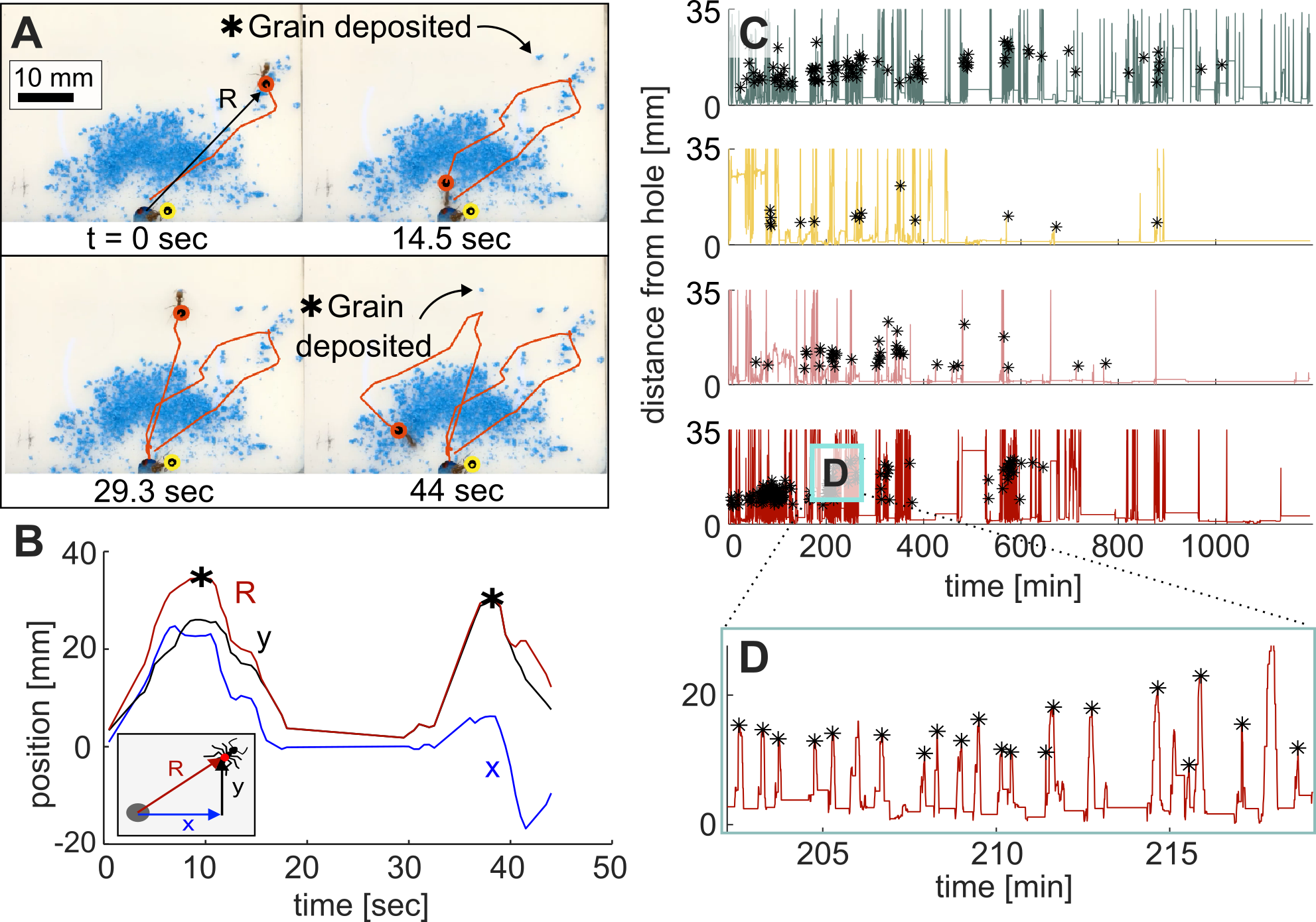}
    \caption{\textbf{Automated high throughput ant tracking method for estimation of grain deposition events}. \textbf{(A)} Lines representing tracked path of a single (red) ant across 44 seconds. \textbf{(B)} The corresponding horizontal position ($x$), vertical position ($y$) and distance from the tunnel entrance over time for the same ant. Asterisks indicate the local maxima in radial distance, which the tracking algorithm denotes as grain depositions. \textbf{(C)} Ant positions, as represented by distance from the tunnel opening, over time for the 4 ants in a single trial. \textbf{(D)} A representation of the position data in (C), for a single ant across $\approx$15 mins. Black asterisks represent points automatically identified as a grain deposition based on the automated thresholding process. }
    \label{fig:track1}
\end{figure*}

Each day, workers were randomly selected from a single colony and their gasters were painted with color codes for purposes of video identification (Figure \ref{fig:exp setup 1}B). New containers were prepared daily, and in each container, we placed 2, 3, 4, 6, 10, 15, 20, or 25 ants. Each trial was started between 4 PM and 6 PM daily, and the containers were randomized such that the same number of ants were not always placed into the same container each day. An array of cameras (Logitech C920) was set up to record overhead videos of each trial for 20 hours. Sample overhead images of pellet depositions over six hours are shown in Figure \ref{fig:exp setup 1}C. Ants were not returned to the colony after each trial.

\subsection{Video analysis techniques}
\label{sec:tracking}

Custom MATLAB scripts were designed to analyze the videos resulting from each trial. For trials with 10 or fewer ants, individual ant locations were tracked via an algorithm which performed color thresholding to identify gaster paint on each ant. Further detail on the identification of ant locations from video data is included in Supplemental Information. A secondary custom MATLAB script converted individual ant positions to radial distances from the tunnel entrance, shown in Figure \ref{fig:track1}A-B. An increase and subsequent decrease in radial distance was presumed to represent an ant leaving the tunnel, depositing a grain, and subsequently returning to the tunnel. The script selected these events by identifying local maxima in radial distance data. From visual observation, we identified a minimum distance from the hole which represented a viable ``trip'' from the tunnel with a grain ($\approx$6 mm), as well as a maximum distance, above which the ant most often exited the camera view and did not return to the tunnel entrance ($\approx$28 mm). Based on visual observation, we also set a maximum time frame for an exit, grain deposition, and return to the tunnel as 12 seconds.  Local maxima which met all of the distance and time criteria described above were counted as grains deposited by that ant. Using this technique, the script tracked each ant's contributions to tunnel construction via the number of grains it deposited over the course of the trial, shown in Figure \ref{fig:track1}C-D. This automated technique was used for ant group sizes of 10 or fewer; for the trials with 15 or more ants, this custom technique was insufficient for tracking multi-color coded individuals, so activity during the first 8 hours of these trials was tracked manually.

\section{Experimental Results}

\begin{figure*}
    \centering
    \includegraphics[scale=0.8]{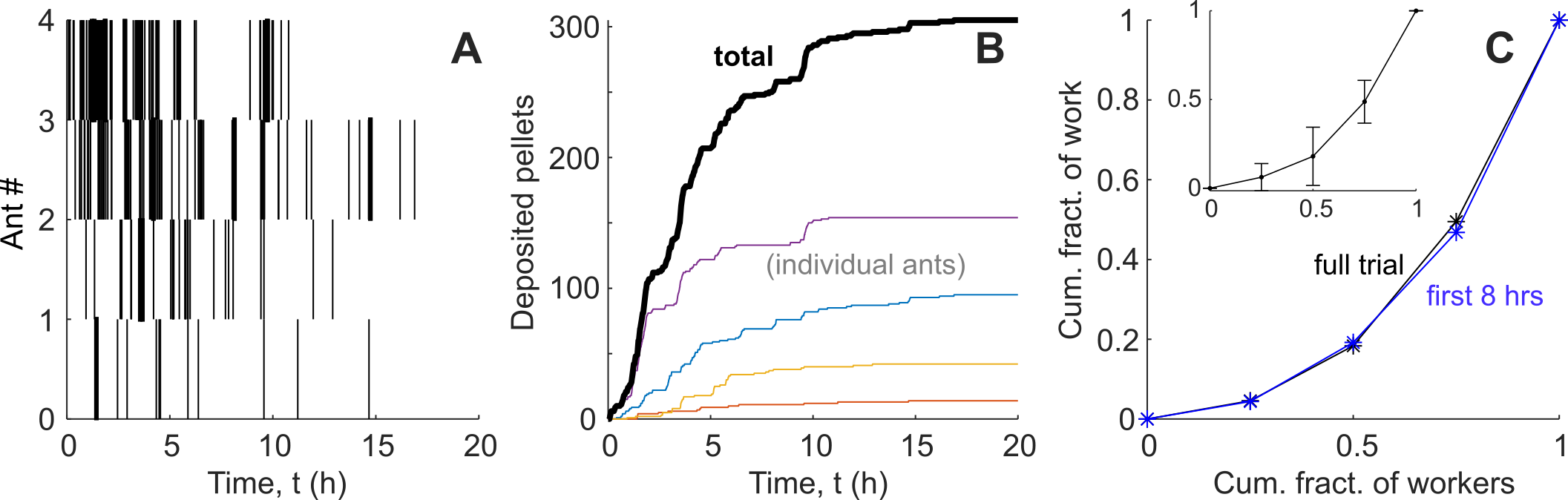}
    \caption{\textbf{Workload inequality is revealed via Lorenz curve analysis of deposition events}. \textbf{(A)} Plots of ant activity over the trial duration, where a black dash indicates a grain transported at that time instance. Each row corresponds to a different ant, with the top row corresponding to the most active ant. Ants are ordered by total activity over 20 hours. \textbf{(B)} Number of deposited pellets, as estimated by the tracking algorithm, for each ant across 20 hours. Colored lines represent individual ant contributions, while the black curve represents the total pellets deposited. \textbf{(C)} Lorenz curves, representing cumulative fraction of grains moved, relative to cumulative fraction of workers involved, for the first 8 hours (blue) and full 20 hour trial (black) for these 4 ants. Inset shows mean Lorenz curve over all 4 trials, with error bars representing standard deviation.}
    \label{fig:exp1}
\end{figure*}

Using the aforementioned video analysis strategy, we denoted the time and location of each ant's grain deposition events over the course of each trial. A sample trial for 4 ants is shown in Figure  \ref{fig:exp1}. The number of deposited pellets increased sharply during the first $\approx$5 hours for each ant. The rates of deposition (and thus, tunnel excavation) decreased drastically afterwards and settled to a slower long timescale behavior. This mirrors results found by Avinery et al. \cite{avinery_agitated_2023} for tunnel growth trends in larger (40 - 70) groups of ants. Similarly, Figure \ref{fig:exp1}A shows the relative level of activity across the trial -- black bars represent active time periods in which grains are transported. The activity started high for multiple ants and decreased over time. This trend is also reflected in the cumulative number of deposited pellets over time (Figure \ref{fig:exp1}B).

\begin{figure}[]
    \centering
    \includegraphics[width=0.6\linewidth]{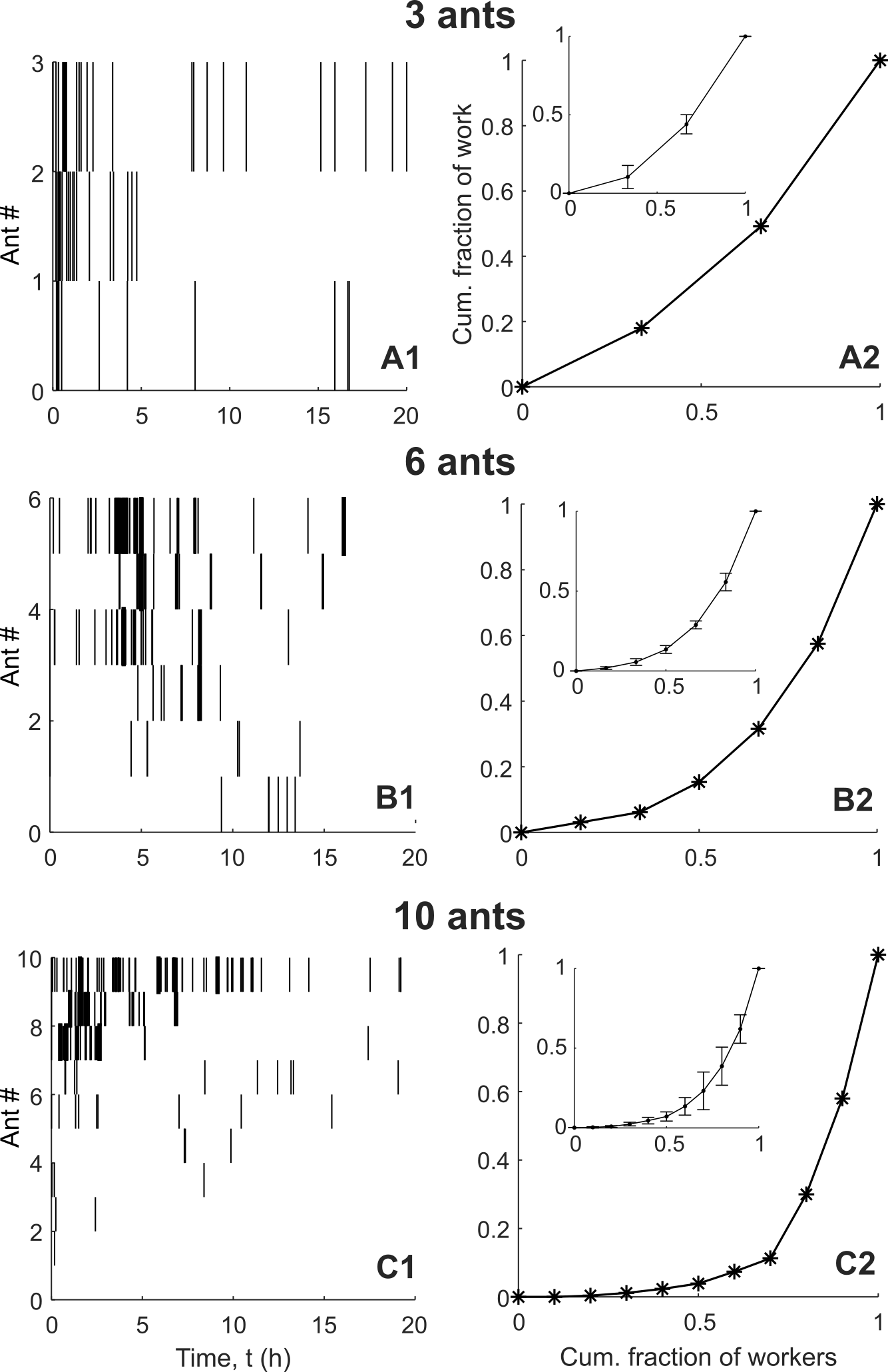}
    \caption{\textbf{Increasing group size leads to greater workload inequality}. \textbf{(A1)} Individual ant activity, as indicated by black dashes, sorted from highest to lowest total activity. \textbf{(A2)} Lorenz curves for the corresponding trial for 3 ants.  Insets show mean Lorenz curves over 4 trials (error bars represent standard deviation). Representative trials are also shown for 6 \textbf{(B1-2)} and 10 \textbf{(C1-2)} ants. }
    \label{fig:exp2} 
\end{figure}

We then investigated how participation, and thus workload, trends evolved throughout these trials. The cumulative fraction of work (approximated by the pellets deposited) completed by a cumulative fraction of ants is shown in Figure \ref{fig:exp1}C. This relationship, also known as a Lorenz curve \cite{aguilar_collective_2018}, represents the level of ``unequalness'' in workload for this particular trial. Perfectly equal workload distribution, or a line of equality, would be represented by a straight line connecting (0,0) to (1,1). Using the experimental Lorenz curves, we calculate the Gini coefficient, or $G$, defined as the ratio of the area between the Lorenz curve and the line of equality to the area under the line of equality. The Gini coefficient represents the deviation of the workload from perfectly equal ($G = 0$, all ants work equally) to completely unequal ($G = 1$, a single ant performs all work). Figure \ref{fig:exp1}C demonstrates that unequal workload distributions are preserved throughout the entire trial.

\begin{figure}[]
    \centering
    \includegraphics[width=0.45\linewidth]{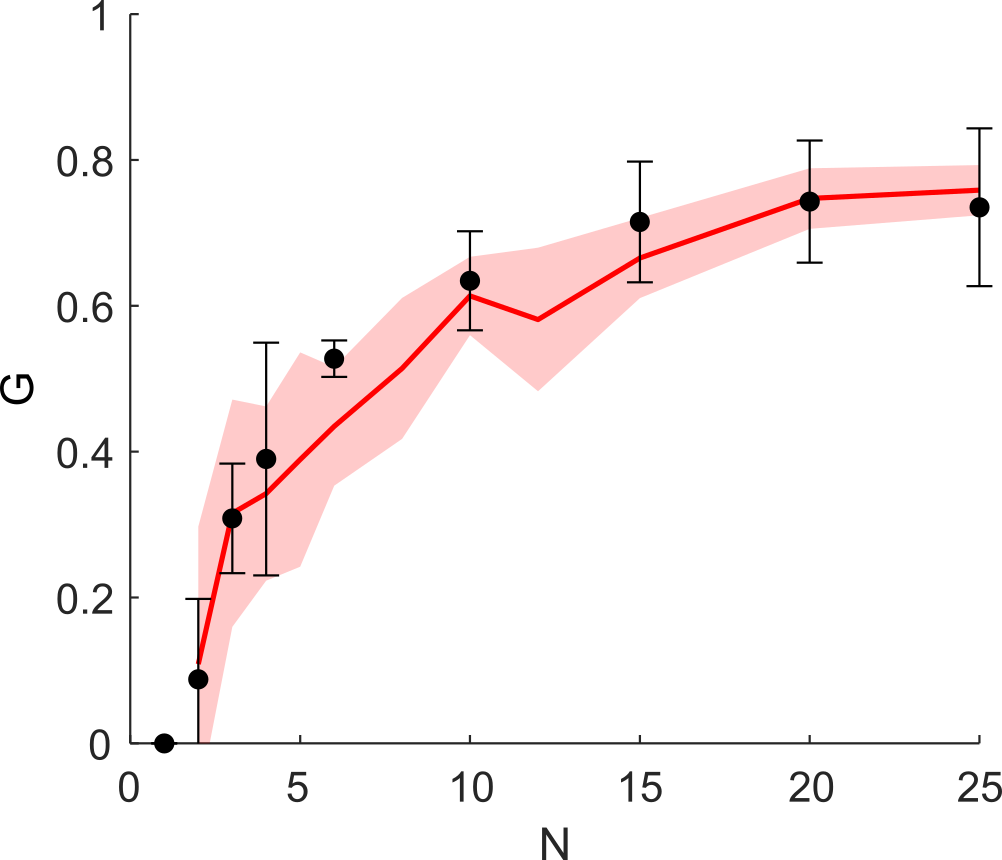}
    \caption{\textbf{Gini coefficient scales sublinearly with group size}. Black dots represent experimental Gini coefficients, $G$, as a function of group size, $N$, derived from Lorenz curves in Figure \ref{fig:exp2}. Error bars represent standard deviations over 4 experimental trials ($<$10 ants) or 3 experimental trials ($\geq$ 10 ants). Red curve represents mean Gini coefficients resulting from the CA simulation for a single choice of parameters ($R$ = 0.1, $\tau_0$ = 10, $C_0$ = 1, $\tau_s$ = 2). The shaded region represents standard deviation over 10 simulation trials performed at each group size. }
    \label{fig:exp3}
\end{figure}

We then compare trends in workload distribution across varying group size. As shown by three representative trials in Figure \ref{fig:exp2}, as the group size increases, the workload distribution becomes increasingly unequal. For the smaller group sizes, nearly all ants participate to some degree (Figure \ref{fig:exp2}A2). However, for larger group sizes, we observe an increasing fraction of ants which perform little to no excavation, and instead a core group of ants perform all digging (Figure \ref{fig:exp2}C2). We then compare trends over all group sizes; the resultant Gini coefficients for the first 8 hours of all trials are represented in Figure \ref{fig:exp3}. We observe a sublinear, monotonically increasing trend in Gini coefficient with increasing group size. If extrapolated, this trend predicts the Gini coefficient range for 30 ants which was predicted by Aguilar et al. ($G = 0.81\pm0.15 - 0.83\pm0.16$) \cite{aguilar_collective_2018}. \footnote{Aguilar et al. used only the ants which appeared in the camera frame at the end of the tunnel to calculate Gini coefficient; thus, their estimates do not take into consideration the inactive workers in the group, while our estimates consider all workers. They report that a fraction of the 30 ants (between $0.22\pm0.1$ and $0.31\pm0.13$) never visited the tunnel face in their experiments, and report a Gini coefficient $G = 0.75\pm0.1$. In our formulation, this corresponds to an effective Gini coefficient range of $G = 0.81\pm0.15$ - $0.83\pm0.16$.}
This trend supports initial observations that large groups not only share work highly unequally, as observed in prior literature \cite{charbonneau_lazy_2015, aguilar_collective_2018}, but that this workload inequality increases the larger the group becomes. 

% To better understand the mechanisms underlying this observation, we implement an agent-based model which captures the role of individual ant decision making in a simplified tunnel environment. 

\section{Cellular Automata Simulation}
\label{sec:CA}
\subsection{CA Model Formulation}
\begin{figure}[]
    \centering
    \includegraphics[width=0.6\linewidth]{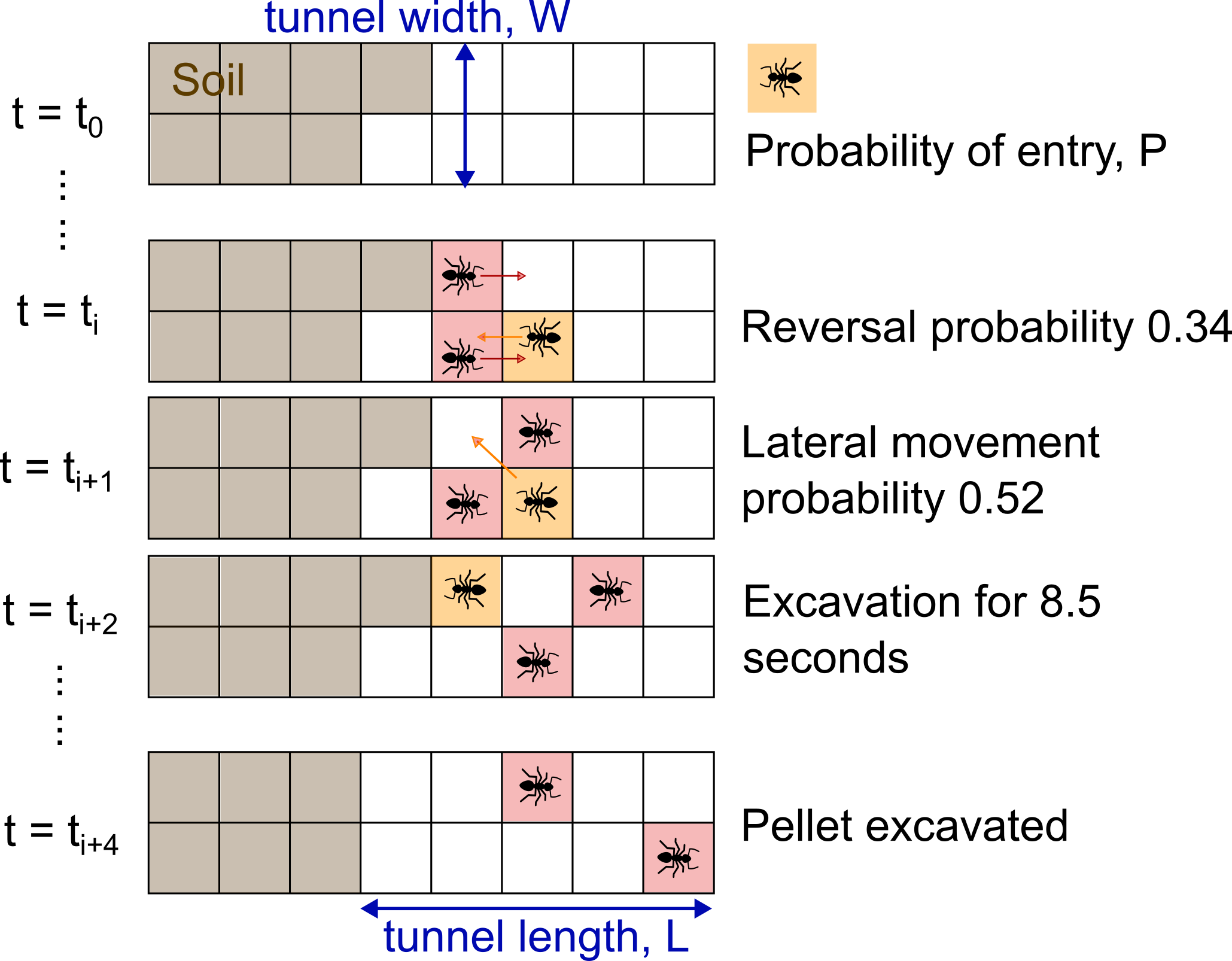}
    \caption{\textbf{Cellular automata simulation layout and algorithm}. Depiction of steps in the excavation of a pellet in the CA simulation, beginning with an ant entering the tunnel with probability $P$, and ending with a pellet excavated and removed from the tunnel. Brown cells represent unexcavated soil in the tunnel, white cells represent already excavated tunnel, orange cells represent ants moving toward the end of the tunnel (left), and red cells represent ants moving towards the tunnel entrance (right).}
    \label{fig:sim_layout}
\end{figure}

To better understand the role of environmental factors in regulating behavior, we build a cellular automata (CA) simulation \cite{aguilar_collective_2018, avinery_agitated_2023}. In this simulation structure, each ``cell'' can be assigned to one of three states: unexcavated substrate in the tunnel, empty (excavated) space in the tunnel, or an ant. The model ants in the simulation move through a series of states as regulated by various factors, shown in Figure \ref{fig:sim_layout}.  
Starting with the ants ``outside'' of the tunnel region, each ant has a probability $P$ of entering the tunnel. Once inside the tunnel, the ants move forward one cell each timestep and have a fixed probability, 0.52, of also moving laterally, as observed experimentally in prior work \cite{aguilar_collective_2018}. If at any point, an ant cannot locomote forward, it either remains in place until space is made, or reverses with probability 0.34 \cite{aguilar_collective_2018}. If the ant reaches the end of the tunnel, it removes a pellet, then exits the tunnel and deposits the pellet. The time required to excavate a pellet, the number of pellets required to remove a CA cell, and the time required to deposit a pellet are all constants which were derived from experimental observation in prior work \cite{aguilar_collective_2018, avinery_agitated_2023, gravish_effects_2012}. After an ant deposits a pellet, the cycle repeats again and the ant has probability $P$ of re-entering the tunnel. Further CA implementation details, including all constants, are described in the Supplemental Information. 

Prior work revealed that during early nest excavation, an ant's probability of entering the tunnel may be modeled by a self-limiting process, which can be effectively described by a ``work-rest imbalance.'' In other words, this imbalance, $t_{imb}$, characterizes how far an ant has deviated from its preferred ratio of work to rest. As described in prior work \cite{avinery_agitated_2023}, an ant's work-rest imbalance can be defined as $t_{imb} = t_{work} - Rt_{rest}$, where $t_{work}$ and $t_{rest}$ represent the amount of time an ant has spent working and resting, respectively. $R$ represents an individual's intrinsic preferred work-rest ratio.  Furthermore, this work-rest imbalance can be used to infer an ant's probability of entering the tunnel and engaging in nest construction. In line with prior work, we use this imbalance to define a probability of entry, $P_0$, which scales with the work-rest imbalance as $P_0 = e^{-t_{imb}/\tau_0}$, where $\tau_0$ a timescale for the tolerance of a work-to-rest imbalance \cite{avinery_agitated_2023}.  According to these definitions, if an ant has rested more than its intrinsic ratio would dictate, it will tend to enter and engage, and in contrast, working more than usual will be more likely to result in resting behavior.

% However, this definition for probability of participation only takes into account an ant's intrinsic willingness to work, $R$, and is not inherently regulated by other potential external factors, such as crowding and confinement. 
% When implementing only $P_0$ as the probability of entry in the CA simulation, assuming all ants have equal $R$, we find that the resultant Gini coefficient is zero over long timescales, for all group sizes tested.  While only some ants participate in digging at any given time, over longer timescales, all ants spend some time both working and resting as dictated by the work-rest ratio, such that net participation over long timescales is nearly equivalent across individuals. This behavior is not representative of trends observed in experimental trials, in which some individuals are consistently less active across long timescales. 

However, this definition for probability of participation only takes into account an ant's intrinsic willingness to work, $R$, and is not inherently regulated by other potential external factors, such as crowding and confinement. When implementing only $P_0$ as the probability of entry, assuming all ants have equal $R$, net participation over long timescales is nearly equivalent across individuals, which is not representative of experimental results.
Instead, we incorporate the role of local crowding information, such as the encounter rate, into individual ants' decision making \cite{aina_toward_2022, avinery_agitated_2023}. 
% Prior studies on robophysical model excavators demonstrated that if robots selectively reduce activity when their local environment is increasingly cluttered, these decisions result in uneven workload distribution which increases the excavation performance  \cite{aina_toward_2022}. 
% Similar to ``override parameters'' introduced in prior work \cite{avinery_agitated_2023}, 
We introduce a term which scales the entry probability by the local success rate of movement in the tunnel. Thus, the probability of entering the tunnel and engaging in nest construction, $P$, can be defined as $P = e^{-1/(lC_0)}P_0$, where $C_0$ is a crowding sensitivity parameter and $l$ is the success rate of moving forward in the tunnel. This success rate is calculated as an exponential moving average with memory $\tau_s$ timesteps. Thus, if an individual ant is entirely unsuccessful at moving forward in the tunnel within $\tau_s$ timesteps, its probability of entry will be zero and it will stop trying to enter the tunnel. In contrast, an ant in an uncrowded environment will be largely unaffected by this additional parameter (its entry probability would be universally scaled by  $e^{-1/C_0}$). 

\subsection{CA Simulation Results}
\begin{figure}
    \centering
    \includegraphics[width=0.6\linewidth]{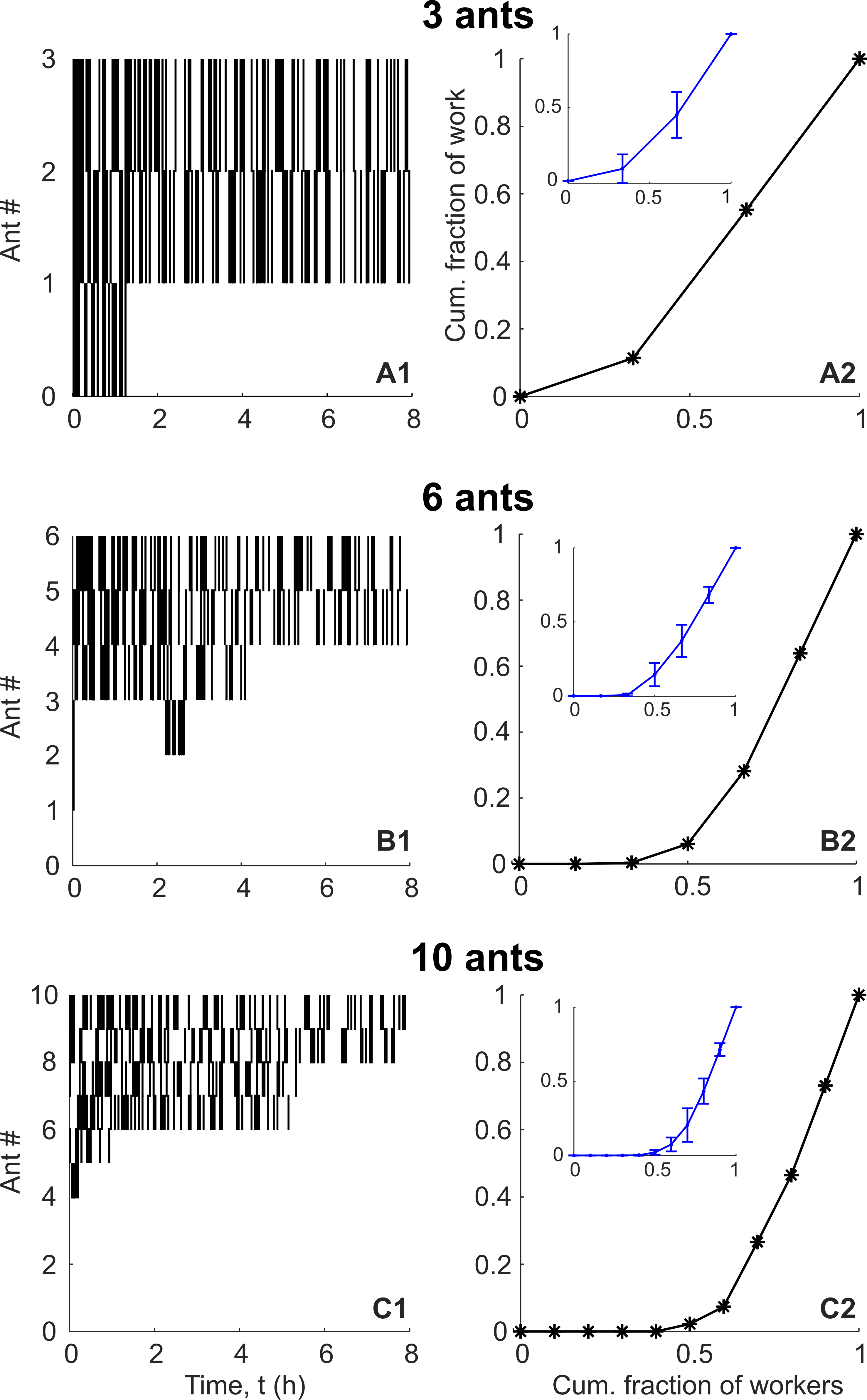} 
    \caption{\textbf{CA Simulation captures correlation between group size and workload inequality}. \textbf{(A1)} Individual ant activity in the CA simulation, as indicated by black dashes, sorted from highest to lowest total activity. \textbf{(A2)} Lorenz curves for the corresponding simulation trial, for 3 ants. Insets represent mean Lorenz curves over 10 simulation runs for one set of simulation parameters ($R$ = 0.1, $\tau_0$ = 10, $C_0$ = 1, $\tau_s$ = 2). Representative trials and mean Lorenz curves are also shown for 6 \textbf{(B1-2)} and 10 \textbf{(C1-2)} ants.}
    \label{fig:sim_results2}
\end{figure}

As shown in Figure \ref{fig:sim_results2}, for one choice of parameters ($R$ = 0.1, $\tau_0$ = 10, $C_0$ = 1, $\tau_s$ = 2), this CA simulation captures overall trends in excavation performance. We observe that in simulation, a group of 3 ants will share workload nearly equally, but in a group of 6 or 10 ants, the work is done by a subset of active workers, and the size of this group does not scale linearly with the total group size. Thus, this simulation effectively demonstrates that the portion of inactive ants will increase with group size.
As shown in Figure \ref{fig:exp3}, for the same choice of parameters, this CA structure is able to capture the experimental relationship between group size, $N$, and workload distribution. However, the variability in simulated Gini coefficient decreases with increasing group size, which is not reflected in the experimental data. 

Overall, the simulated ant activity is less stochastic than that observed in experiments: in the simulation, ants which are active tend to excavate throughout the entire trial, and those which stop excavating tend to have little no activity for long durations. This tendency results in Lorenz curves for the simulation which are largely comprised of straight line segments (Figure \ref{fig:sim_results2}) -- there is a long flat ``tail'' comprised of inactive or nearly inactive ants, and a diagonal straight line segment which represents the active ants, which often perform similar levels of work. 
This tendency to continuously engage or disengage from excavation is likely a byproduct of the short ``memory'' of ants in simulation, dictated by the memory timescale $\tau_s =2$. In other words, a short memory means that ants are quick to decide whether or not to engage in digging based on local crowding conditions, and these decisions persist throughout the trial. 

\subsection{CA Parameter Sensitivity}
\begin{figure*}
    \centering
    \includegraphics[scale=0.75]{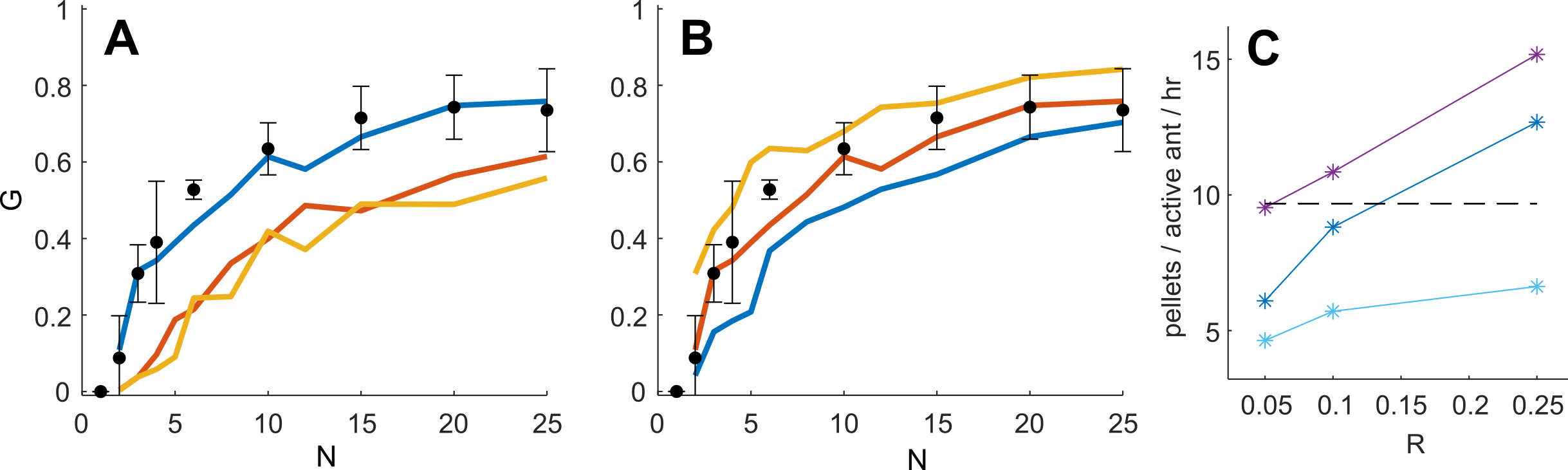}
    \caption{\textbf{Effect of simulation parameters on workload distribution and excavation rates}. \textbf{(A)} The effect of memory parameter $\tau_s$ on resulting mean Gini coefficient, across all group sizes, N, in simulation (colored lines). Blue, red, and yellow curves represent $\tau_s$ = 2, 5, 10, respectively. Experimental measurements are shown with black dots and error bars. \textbf{(B)} The effect of intrinsic work-rest ratio, $R$, on mean Gini coefficient trends. Blue, red, and yellow curves represent $R$ = 0.05, 0.1, 0.25, respectively. \textbf{(C)} The effect of simulation parameters on excavation rates (colored lines) vs. mean experimental excavation rate (dashed black line). From bottom to top, colored curves represent $\tau_0$ \& $C_0$ = 1, 10, 100, respectively.}
    \label{fig:sim3}
\end{figure*}

We next seek to understand how sensitive the simulation results are to changes in individual parameters and probe the role of individual mechanisms in emergent collective behavior. We run a set of simulations in which we systematically sweep over each simulation parameter individually and otherwise use a set of default parameters ($R$ = 0.1, $\tau_0$ = 10, $C_0$ = 1, $\tau_s$ = 2).
The results of this parametric sweep are shown in Figure \ref{fig:sim3}. We observe that the Gini coefficient trends are largely insensitive to the crowding sensitivity parameter $C_0$, as well as the timescale for the tolerance of a work-rest imbalance, $\tau_0$. The simulation results for sweeps over these two parameters are contained in the Supplemental Information. However, we do observe a dependence of Gini coefficient on memory parameter $\tau_s$, as well as the intrinsic work-rest ratio, $R$ .  We observe that in order to effectively reproduce experimental trends in Gini coefficient, a short timescale ($\tau_s$ = 2) is necessary, corresponding to a memory of approximately 7.4 seconds based on the simulation timestep used in this work (Figure \ref{fig:sim3}A). We hypothesize that in order to maintain an appropriate rate of crowding-induced deactivation, the ants in simulation are effectively ``impatient,'' persisting only for short periods in crowded conditions before their probability of re-entry diminishes. 
We also observe that for higher intrinsic motivation, $R$, we observe increasingly unequal workload distribution (Figure \ref{fig:sim3}B). We hypothesize that if ants have a higher baseline likelihood of entering the tunnel (higher $R$), the tunnel will become more crowded, leading to increased levels of crowding-induced ``deactivation.'' In contrast, if ants have a low baseline motivation, there will be less ants entering the tunnel and less crowding. As a result, more ants are able to participate in excavation throughout the trial without persistent roadblocks. 

We also investigate how the parameters introduced in this simulation affect excavation rates. Prior work has shown that total excavated tunnel area (and by association, the number of excavated pellets) scales linearly with the total number of active ants \cite{gravish_effects_2012}. Thus, we compute the total number of active ants in each hour of the simulation for each combination of parameters. We then compare the mean number of active ants to the total number of pellets excavated in each trial and fit a linear excavation rate, measured in pellets excavated per ant per hour.  Figure \ref{fig:sim3}C shows the excavation rates predicted by the simulation for three combinations of parameters and shows how the ants' intrinsic motivation, $R$, is strongly correlated with resulting excavation output. 
% Additionally, the combination of tolerance timescale $\tau_0$ and crowding sensitivity $C_0$ influences rates, with increasing parameter values boosting overall activity. 
Overall, we observe that for all parameters tested, the simulation excavation rates are within $\pm50$\% of experimental values. The role of each individual parameter on excavation rates, as well as details of how excavation rates were approximated, are described further in the Supplemental Information.

\section{Theoretical Analysis}

\begin{figure}[]
    \centering
    \includegraphics[width=0.65\linewidth]{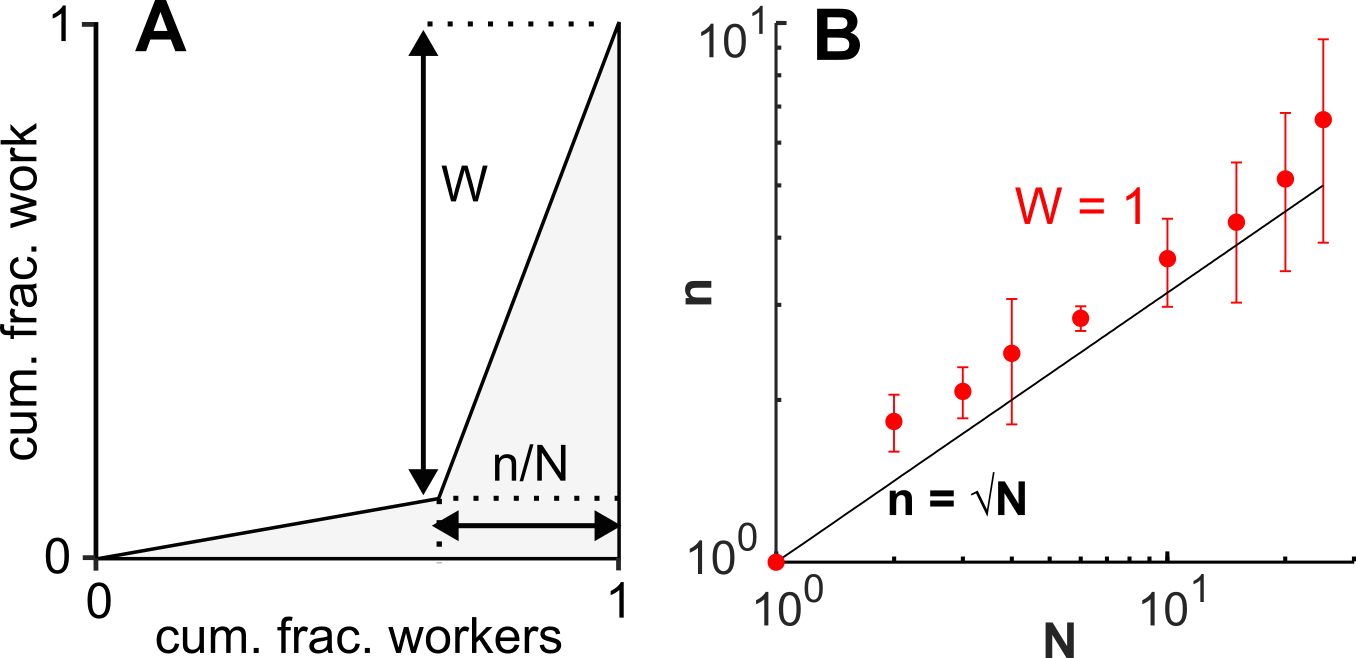}
    \caption{\textbf{Participation trends are captured by a square root scaling relation}. \textbf{(A)} A simplified version of the Lorenz curve in which we assume that a subset of active ants, $n$, contribute equally to the excavation, performing fraction $W$ of the total work. \textbf{(B)} Using a simplified form for approximating Gini coefficient from the Lorenz curve (shown in A), we estimate the total number of active ants from the measured Gini coefficients, assuming $W$ = 1.  Red points with error bars represent means and standard deviation over 4 trials (2-6 ants) and over 3 trials (10 - 25 ants). Solid line represents the case where $n$ scales as the square root of $N$. }
    \label{fig:discussion1}
\end{figure}

Both the experimental and simulation results demonstrate a consistent, sublinear relationship between group size and Gini coefficient. To further understand the origin of this trend, we approximate a relationship between Gini coefficient and net participation in a group. Given prior observation that some subset of ants within a group is active while the rest remain primarily inactive, we choose to create a simplified definition for Gini coefficient. Shown in Figure \ref{fig:discussion1}A, we assume that a typical Lorenz curve can be reduced to two straight line segments: one which captures the low activity in the inactive subset of ants, and a second, steeper line segment which captures the majority of work completed by the ``active'' ants. Thus, we assume that some subset, $n$, of active ants, completes portion $W$ of the total work done. With this simplification of the Lorenz curve, we can directly relate Gini coefficient ($G$) to the number of active ants ($n$) by: 

\begin{equation}
{G = W - \frac{n}{N} }
\end{equation}

Using this relation, we then convert experimentally derived Gini coefficients to an estimated number of active ants for each group size. We assume that the active ants complete nearly all of the total work ($W$ = 1). When comparing the approximated number of active ants, $n$, to group size, $N$ (Figure \ref{fig:discussion1}B), we observe that the number of active ants scales as the square root of the total group size, or that $n \propto \sqrt{N}$.

\subsection{Analytical approach to model participation rates}
To identify a physical mechanism and biological motivation for this square root scaling, we develop a simplified model for ant participation in digging using a rate equation methodology \cite{murray_mathbiology}. This model simplifies the cellular automata model by abstracting ant movement within the tunnel as a stochastic process. We introduce a model tunnel environment which, similar to the cellular automata, is 2 cells wide and $L$ cells long. We assume that the number of ants in the tunnel at any given time, $n$, corresponds to the number of active ants in the collective, and that the tunnel is sparsely filled $(n\ll L)$. The total number of ants in the model is $N$, some of which are within the tunnel (active) and some of which are outside the tunnel (inactive). 

We define the rate of change in the number of ants in the tunnel as the number of ants entering minus the number of ants leaving. We assume that a fraction ($c_1$) of $N$ total ants are trying to enter the tunnel of fixed length $L$ at any given time. We also assume that ants leave the tunnel at a rate proportional to the probability of finding the tunnel blocked in a given traversal. This assumption corresponds to the crowding-induced inactivity which is incorporated into the cellular automata. Thus, we define the rate of change of active ants in the system as: 

\begin{equation}
\deriv{n}{t} = c_1N - c_2 P_f
\label{eqn: rate}
\end{equation}

where $c_1$ and $c_2$ are constants, and $P_f$ is the probability that an ant encounters a blockage during tunnel traversal (and thus, denotes a failed pellet excavation). We assume that the tunnel is excavated very slowly relative to ant movements, and thus the tunnel length, $L$, can be treated as constant at any point in time. This simplification also ignores the time delay between an ant choosing to leave the tunnel and doing so.

An ant successfully traverses the tunnel if it is able to reach the other side without running into a blockade. Because the tunnel is two body widths wide, we can say that the tunnel is blocked at a specific location if there are two ants across the width of the tunnel at that location (as shown in Fig. \ref{fig:sim_layout} at time $t_i$). We assume an area of $2L$ (twice the tunnel length) with $n$ active ants randomly distributed throughout the tunnel. We find that the probability of a blockage occurring at any given location in the tunnel, at a single instance in time, scales quadratically with the density of ants in the tunnel:
\begin{equation}
{P_b = \frac{n(n-1)}{2L(2L-1)} \approx \frac{n^2}{4L^2}}
\end{equation}
As an ant moves through the tunnel to excavate a grain and then deposit, it must move a total distance $2L$, with a probability of finding a blockade at every step. Because other ants also move every step, we assume that the probability of a blockade is history independent (ants randomly redistribute each step). This simplification ignores ant interactions - ants moving systematically could affect these probabilities conditionally. With this assumption, the probability of successfully traversing the tunnel is:
\eqn{P_s = (1 - P_b)^{2L} \approx 1 - 2LP_b \approx 1 - \frac{n^2}{2L}}
The above uses the binomial approximation, which assumes $\frac{n^2}{L^2} \ll 1$. Therefore, the probability of an ant experiencing a blockage at any point during a tunnel traversal is:
\eqn{P_f = 2LP_b \approx\frac{n^2}{2L}} 

Substituting this into Equation \ref{eqn: rate}, we find:
\eqn{\deriv{n}{t} = c_1N - c_2\frac{n^2}{2L}}
Here, we treat the length of the tunnel as roughly constant relative to the changes in the numbers of ants and fold $2L$ into $c_2$. We note that in experiment, we observe an increase in activity at the onset of digging, followed by a decay to a steady state over long times (see Figures \ref{fig:exp1}-\ref{fig:exp2}). At the peak of activity and after a long time, we observe that $\deriv{n}{t} = 0$. Using this condition, we find that the model predicts experimental scaling observations:

\begin{equation}
n \propto \sqrt{N}
\label{eqn: sqrt scaling}
\end{equation}

\subsection{Optimality of excavation rates}
We posit that ants attempt to maximize the speed of digging. The rate of digging is proportional the number of ants traversing the tunnel, which is equal to the number of active ants multiplied by their probability of successful traversal:
\eqn{D = n P_s = n\Big (1 - \frac{n^2}{2L}\Big )}
Where $D$ is the rate of deposits and $P_s$ is the probability of a successful traversal. To find the number of active ants, $n$, that would maximize $D$, we differentiate $D$ with respect to $n$ and find its roots:
\eqn{\deriv{D}{n} = 1 - c\frac{n^2}{2L} = 0}
Evaluating, we find:
\eqn{n \propto \sqrt{L}}
The number of active ants required to maximize digging is proportional to $\sqrt{L}$. From Equation \ref{eqn: sqrt scaling} and experimental results, we observe that the number of active ants scales with $\sqrt{N}$. This means that the ants would maximize digging rates in a scenario where the length of the tunnel scaled with the total number of ants, or
\eqn{N \propto L} 
In other words, ants' observed behaviors would optimize digging speeds if tunnel length was directly proportional to the number of ants. Observing this scaling law directly in experiment would be difficult. As ants dig tunnels, the lengths of the tunnels change, which in turn impacts the optimal group size. Additionally, large numbers of ants do not dig a single tunnel - instead, ants are distributed across a nest, digging tunnels simultaneously. In these scenarios, defining the group size for a single tunnel becomes difficult. Nevertheless, it has been observed that the size of a group of ants is positively correlated with both the size and complexity of an ant nest and the length and speed of digging \cite{gravish_effects_2012, nestexca2001, buhl2004nest}. Also, a smaller group of ants will stop digging a tunnel earlier than a larger group \cite{bruce2019digging}. Therefore, in natural environments where ants have already begun to dig tunnels, ants that are part of larger colonies would naturally be digging in larger tunnels. If this correlation is linear, the behavior observed in this study would indeed optimize for digging speed.

\section{Discussion}

Collectives often utilize division of labor to accomplish goals in constrained environments. Here, we analyze one collective system, tunnel construction in \textit{S invicta} fire ants, through experiments, simulation, and theoretical modeling. We observe the emergence of unequal workloads as numbers of workers rise, as assayed by Lorenz curves and corresponding Gini coefficients. Previous studies which analyzed fixed numbers of ants were unable to identify a mechanism for workload inequality. By systematically varying the number of ants in a fixed experimental area, we find that the correlation between group size and inequality can be described by a simple scaling law: the number of active ants is approximately the square root of the group size.  This finding parallels scaling laws observed in other social systems which previously lacked mechanistic explanations. This result also rationalizes previous observations of workload inequality by Aguilar et al. \cite{aguilar_collective_2018} and provides an explanation for observed behavior.
Our CA simulation and theoretical model both suggest that local cues, such as the levels of confinement or encounter rate, may dictate individual decision making and thus underlie emergent workload inequality.  This result mirrors prior observation that encounter rate may act as a cue for local density, and in turn help regulate other aspects of social insect behavior \cite{gordon2021movement, gordon2010ant}.
In this work, we suggest that if individuals choose to remain inactive when encounter rate is high, a quadratic failure rate emerges which results in a square root scaling in participation rate. This result represents one of the first mechanistic explanations for the relationship between workload inequality and group size in collective systems. 

Several scaling laws from other domains suggest that in large groups, only a small fraction of the group is responsible for the majority of productive output. For example, in the late 19th century, Pareto observed that in economics, approximately 80\% of wealth was concentrated in about 20\% of a population, becoming the basis for the modern ``Pareto Principle'' \cite{sanders1987pareto}. Similarly, Price showed that this extreme inequality increased with group size; in observations of scientific publishing, Price demonstrated that roughly the square root of the total authors were responsible for one half of all published papers \cite{price1963little}. Zipf showed that in the written language, a word's frequency is inversely proportional to its rank \cite{zipf1932selected}, an observation which extends to city populations and income distributions \cite{hill1974rank, wyllys1981empirical}.  
% The Ringelmann effect has been used to describe scenarios in which an individual's contribution to a task decreases with increasing group size \cite{ingham1974ringelmann}.

However, these scaling laws are largely observational and cannot point to a specific causal mechanism. The principles underlying scaling laws for inequality, in both human societies and other biological collectives, remain unclear. In this study, we provide an explanation involving the rate of blockages in the tunnel - this corresponds to a failure rate proportional to $n^2$, which results in the square root scaling. Interestingly, particle collisions in a fixed volume also scale quadratically with population density. This may point towards similar scaling laws in other constrained biological collectives where collisions strongly influence dynamics, such as within honeybee nests or ant rafts. However, we believe that these trends may not hold for \textit{S. invicta} ants at higher populations - at colony sizes, ants would not only dig one tunnel. Instead, we expect that ants would dig many tunnels in parallel, as we observe in nature. Further work is needed to understand how these patterns of workload inequality emerge in tasks which are differently constrained.  

Importantly, our modeling shows that these global changes can emerge via local interactions: the quadratic failure rate in our theoretical model emerges if individuals choose to exit the tunnel when they are ``blocked.'' Similarly, in the CA model, individuals make decisions based on their rate of successful movement in the tunnel. Only when this individual decision making is introduced into the CA is the simulation able to accurately capture the square root scaling relationship. Thus, both models suggest that local information drives individual decision making and results in rates of emergent inequality observed in the collective. This process bears resemblance to response threshold models, which suggest that workers are active when external stimuli exceed internal thresholds \cite{jeanson_emergence_2007, bonabeau_fixed_1998, robinson2019genetic}. Such models have been used to predict the behavior of other active matter systems besides social insects \cite{lavergne2019group}. The CA model explored in this work bears some similarity to variable response threshold models, in that an ant's intrinsic work-rest ratio is fixed, but the rate of engagement with the task is modulated by external factors.

Workload inequality may serve a variety of adaptive functions in overall collective success; for example, using CA simulations in a tunnel environment, Aguilar et al. suggest that $G\approx0.75$ corresponds with optimal traffic flow and fastest digging speeds (using their formulation for Gini coefficient). However, this Gini coefficient was optimal only for a group size of 30 ants: another CA simulation with 60 ants showed that the optimal Gini coefficient increased to $\approx$0.8. The present study echoes the idea that optimal participation rates likely vary with the group size. In our theoretical analysis, we suggest that the changes in participation rates observed (namely, the square root scaling) would optimize digging speeds if the group size was proportional to the length of the tunnels. This analysis provides a path towards understanding how decision making at the individual level can allow a group to maintain high performance in the presence of variable environmental constraints.

However, our mechanistic analysis relies on the assumption that spatial constraints are the primary driver of group dynamics.  In tasks with different constraints, many other explanations are possible to describe the adaptive function of workload inequality \cite{dornhaus2008not}.  Some subset of studies argue that maintaining a ``reserve'' of inactive ants allows the colony to continue to function when active workers stop working \cite{charbonneau_who_2017}, increasing the long-term persistence of the colony \cite{hasegawa_lazy_2016}. Others suggest that the colony produces ``extra'' workers, choosing only to employ the most efficient workers of a given group to optimize performance \cite{bernadou2024randomness}.
Some argue that under certain conditions, inequality may simply be a byproduct of the task allocation process, without necessarily serving an adaptive function \cite{khajehnejad_explaining_2023}. Future studies may seek to explore the functional advantages of workload inequality across a variety of tasks within a colony. 
Certainly, under the presence of environmental constraints, choosing to employ the optimal subset of individuals may prevent deleterious clogs and traffic jams, ultimately aiding collective success.

\section{Materials \& Methods}

\subsection{Details of custom MATLAB algorithm for ant tracking}

For trials with 10 or fewer ants, individual ant locations were tracked via an algorithm in MATLAB which performed color thresholding and isolated connected components, or groups of adjacent pixels. The largest connected components within each pre-identified color range (corresponding to color markings on ant gasters) were isolated. For each connected component, we computed both the size of the connected region, and the distance of each connected region to its location in the prior frame. We computed a weighted sum of these two parameters (size of region and distance to prior location). The weights were chosen via manual trial and error to maximize tracking accuracy. 
For each color, we then chose the connected region with the maximum weighted sum.  These regions were presumed to belong to individual ants of each color code and were used to assign the pixel location of each ant. If a particular ant was not identified within the camera region, its location was assigned as either inside the tunnel, or at the camera boundary if the ant was near a boundary in a prior frame. 

This tracking methodology allowed for automatic estimation of total workload via counting trips to/from an excavation site. The MATLAB algorithm was capable of analyzing 20 hours of video data, and subsequently calculating workload and activity information, in several minutes.  In prior work, individual contributions had often been tracked manually, requiring additional labor \cite{aguilar_collective_2018, avinery_agitated_2023}.  While we focus specifically on fire ant digging in this paper, this tracking method could be employed to study workload distributions in any social insect collective for which a task involves repeated trips to a singular site. The method introduced in this study represents a major step towards more rapid monitoring of workload inequality in social insects. However, this method remains untested for groups larger than 10 ants-- in these cases, we anticipate that using more advanced software may be required to accurately monitor behavior. 

\subsection{Details of CA simulation}

The cellular automata simulation used in this work was built in MATLAB and was run using MATLAB's Parallel Computing Toolbox to accelerate simulation times. The simulation relied on an agent-based model based on prior work by William Savoie \cite{aguilar_collective_2018} that was built upon a tunnel whose cells can take on various states: unexcavated (filled with sand), excavated, empty (no sand), or containing a single ant. Each ant in the simulation can have one of five activity states: moving in tunnel (ascending), moving in tunnel (descending), digging, depositing a pellet, or resting (waiting for $P > e^{-1/(lC_0)}P_0$). The simulated ``tunnel'' is comprised of cells which are one ant body width wide and one ant body width long (accomodating a single ant at any given time). The simulated tunnel is two cells wide and 100 cells long. At the simulation times explored in this work, this length is sufficient to guarantee that no group of simulated ants will dig the full length of the tunnel. We assume that ants travel one cell in either direction per timestep. Using estimations of ant walking speed from prior literature as 0.27 body lengths per second \cite{avinery_agitated_2023}, we approximate the real-world time of a single simulation timestep as 3.7 seconds (equivalent to the time it takes an ant to travel a single body length). Ants in the tunnel move laterally with probability 0.52 and reverse with a probability 0.34 when blocked by ants below them.

We utilize results from prior literature to estimate the number of ant trips which resulted in a single cell being removed from the end of the tunnel (representing tunnel excavation). We found that it takes 206 trips per cm of excavation \cite{avinery_agitated_2023}, or 40 trips per cell excavated, assuming two cells must be excavated for a full row of cells to be removed from the tunnel. Similarly, observations from prior work revealed that ants spend an average of 8.5 seconds excavating \cite{avinery_agitated_2023}, which we approximate as 2 simulation timesteps. All parameters used in the CA simulation and additional results from a single CA simulation are shown in the Supplemental Information.

\subsection{Estimating excavation rates across all experimental trials}

To estimate excavation rates from all experimental data, we assume that excavation rates scale roughly linearly with the number of ants which engage in digging, as observed in \cite{gravish_effects_2012}. We record the number of pellets transported during each hour of every experimental trial. We also record the number of ants who transport any number of pellets during each of those hour-long periods, denoting this quantity as the number of ``active ants'' in each period. We then compare pellets deposited in each hour vs. the number of active ants in each hour; the hourly excavation rates across experimental trials are highly variable.  We then fit a line to this data via least squares regression. The resultant fit rate is $\approx$9.5  pellets per active ant per hour of digging. This excavation rate represents a rough proxy for the number of pellets excavated by a single ant in one hour. Plots of all experimental excavation rates, and fit lines, are shown in the Supplemental Information.

\subsection{Calculating excavation rates in CA simulation data}

To estimate excavation rates for simulation trials, we use a similar procedure to the one used to calculate experimental excavation rates. For all 10 iterations run at each set of simulation parameters, we record the number of pellets transported during each hour. We also record the number of ants that transport at least one pellet during each of those hour-long periods. This results in a total of 880 hours for each set of parameters (8 hours in each simulation x 10 iterations x 11 different group sizes tested for each set of parameters). Fitting a line via least squares regression to this data for a single set of parameters ($C_0$ = 10, $R$ = 0.1, $\tau_0$ = 10, $\tau_s$ = 2) yields a resultant fit rate of $\approx$8.1  pellets per active ant per hour. We also analyze the rates if the total number of pellets transported at the end of each 8 hour trial is used for calculation instead. The excavation rate estimated from a linear fit to these 110 trials is the same as that for all hours of simulation (for the same set of simulation parameters), and thus we use this simplified form of the calculation for reporting simulation excavation rates for each set of parameters. All plots of simulated excavation rates are shown in the Supplemental Information.

\newpage

\section{Acknowledgments}
We thank William Savoie for providing template code from his prior work \cite{aguilar_collective_2018}, upon which we based the CA simulation used in this work. We also thank Ram Avinery for insightful guidance based on his prior work in CA modeling, particularly his model which similarly relied on an intrinsic work-rest ratio. 
We also thank various members of the Goodisman lab: namely, Paige Caine and Andrew Robertson, who assisted with ant husbandry throughout the time frame that the experiments were conducted. We also thank Prof. Takao Sasaki for providing advice on labeling and painting strategies for fire ants. \\

\noindent\textbf{Funding:} 
National Science Foundation Grant \#NSF-IOS-2019799. \\

\noindent\textbf{Author Contributions:}
Conceptualization: DIG, MADG, LKT, AR
Methodology: LKT, AR
Investigation: LKT, NW, AR
Data curation: NN, NW, LKT
Supervision: DIG, MADG
Writing-original draft: LKT, AR
Writing- review \& editing: LKT, AR, DIG, MADG, NN, NW \\

\noindent\textbf{Data and materials availability:} All data are available in the main text or the supplementary materials. Additional code and/or data are available from the authors upon request. 

\newpage
\section{References}
\bibliography{ant_ref}

\newpage

\section{Supplemental Information}

\subsection*{CA simulation}
\begin{figure}[h]
    \centering
    \includegraphics[width=0.9\linewidth]{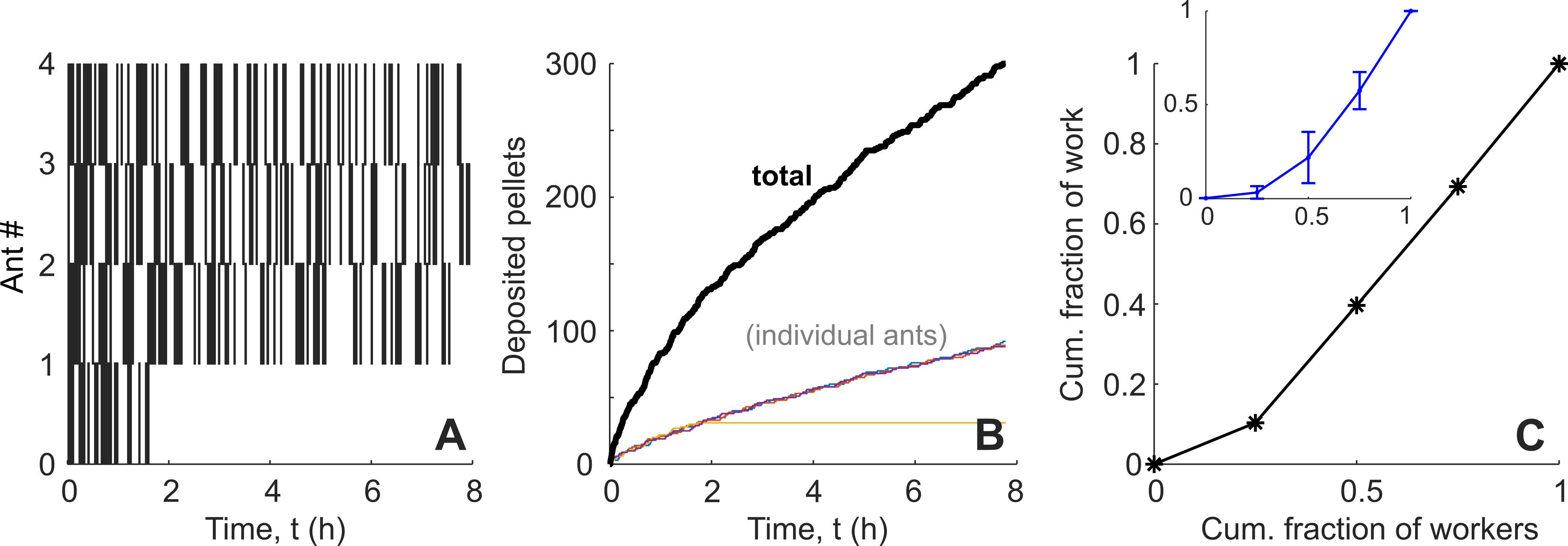}
    \caption{\textbf{Single 4 ant trial simulation results}. \textbf{(A)} Plots of ant activity over the simulation duration, where a black dash indicates a grain transported at that time instance. Ants are sorted from highest to lowest activity (top to bottom). \textbf{(B)} Number of deposited pellets for each simulated ant, across the simulation duration. Colored lines represent individual ant contributions, while the black curve represents the total pellets deposited. \textbf{(C)} Lorenz curve, calculated from deposited pellets. Inset represents mean Lorenz curves across all 10 simulation trials for one set of simulation parameters ($R$ = 0.1, $\tau_0$ = 10, $C_0$ = 1, $\tau_s$ = 2). }
    \label{fig:sim_results1}
\end{figure}

All parameters used in the CA simulation are shown in Table \ref{fig:sim_table}. Results from a single CA simulation trial, including ant activity, excavation rates, and Lorenz curves, are shown in Figure \ref{fig:sim_results1}.

\subsection*{Estimating excavation rates across all experimental trials}

\begin{figure}
    \centering
    \includegraphics[width=0.85\linewidth]{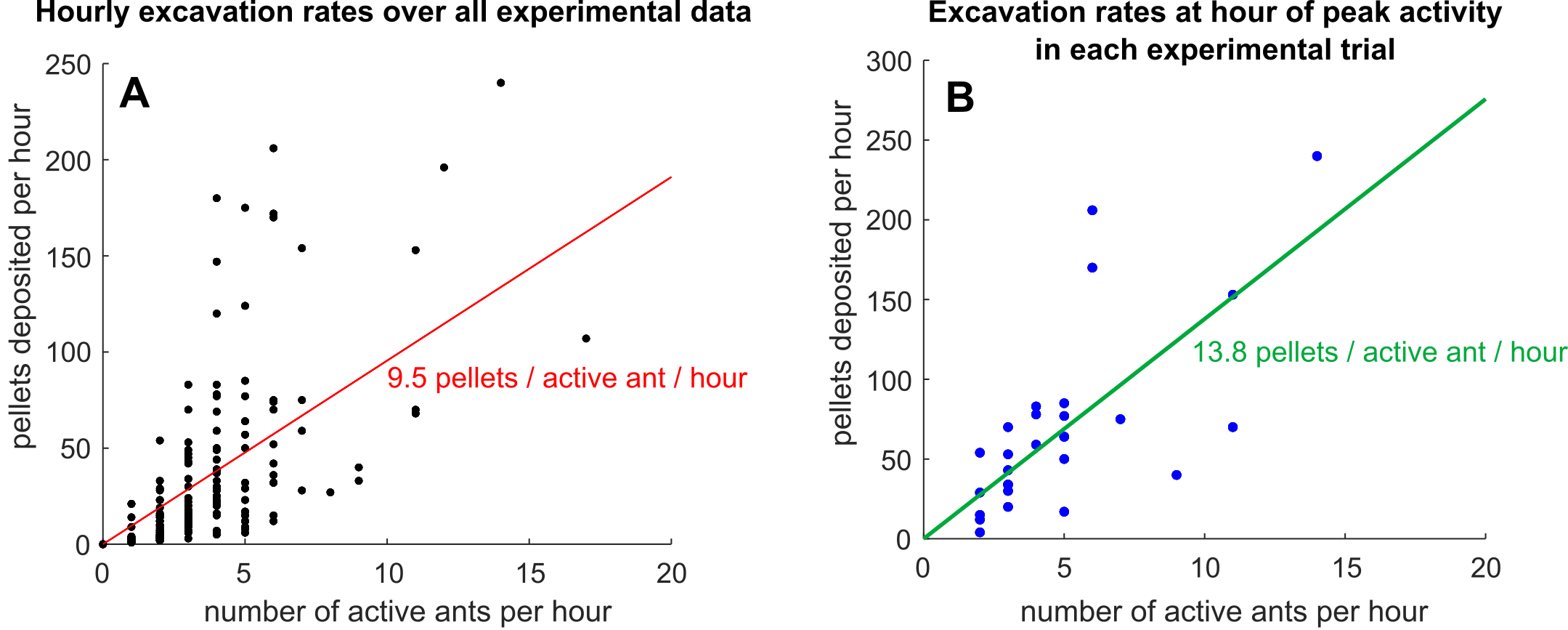}
    \caption{\textbf{Method for calculating excavation rates in experimental trials} \textbf{(A)} The number of pellets deposited in each hour of experiment are plotted vs. the number of active ants in each hour of experiment. Red curve represents the line of best fit, with an effective mean excavation rate of 9.5 pellets/active ant/hour. \textbf{(B)} The number of pellets deposited in the most active hour of experiment are plotted vs. the number of active ants in each of these hours. The green curve represents the line of best fit, with an effective mean maximum excavation rate of 13.8 pellets/active ant/hour.}
    \label{fig:SI_exp_rates}
\end{figure}

We plot pellets deposited in each hour of experiment vs. the number of active ants in each hour in Figure \ref{fig:SI_exp_rates}A.  We then fit a line to this data via least squares regression; the resultant fit rate is $\approx$9.5  pellets per active ant per hour of digging (shown in Figure \ref{fig:SI_exp_rates}A).  For comparison, we examine the hour in which the most pellets were excavated in each trial. These data points are represented in Figure \ref{fig:SI_exp_rates}B. In this case, the excavation rate for the hours of maximum activity is 13.8 pellets per active ant per hour.

\subsection*{Calculating excavation rates in CA simulation data}
\begin{figure}
    \centering
    \includegraphics[width=0.85\linewidth]{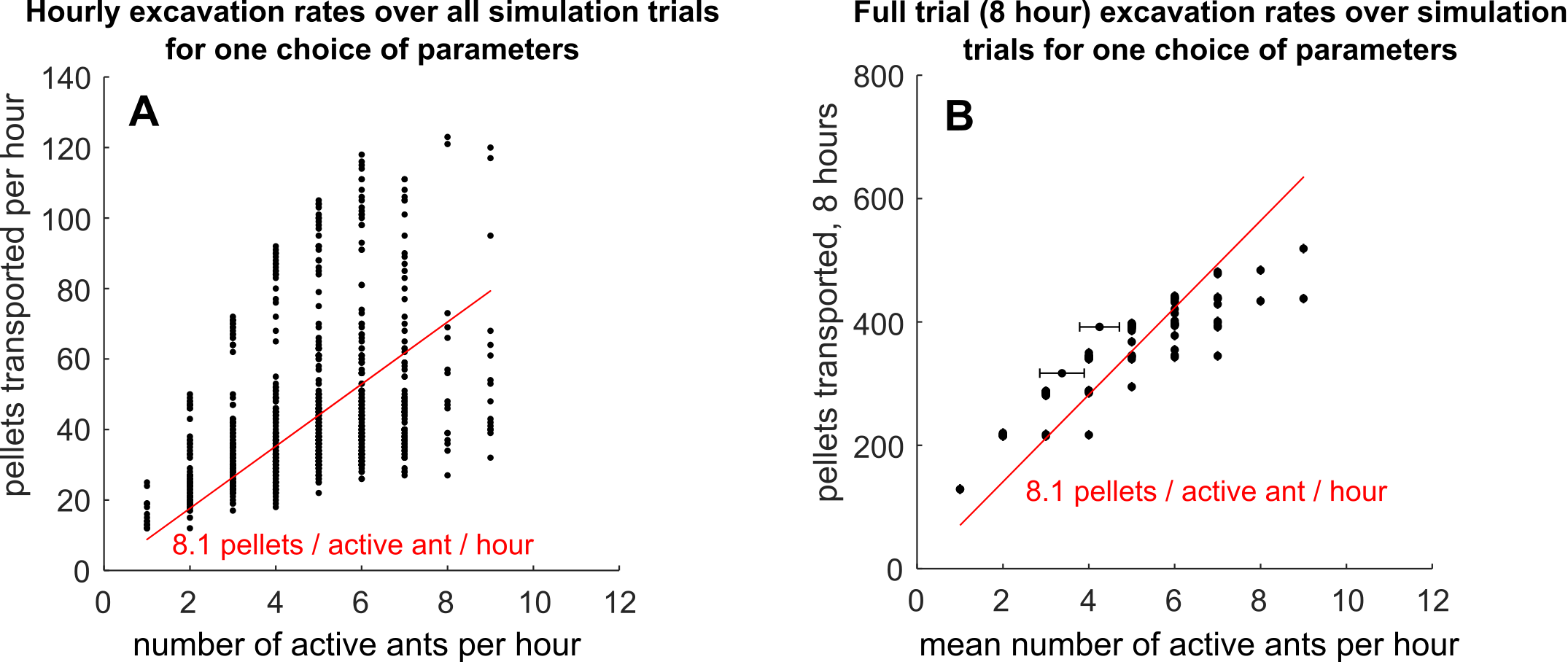}
    \caption{\textbf{Method for calculating excavation rates in simulation trials} \textbf{(A)} The number of pellets deposited in each hour of simulation at one parameter combination ($C_0$ = 10, $R$ = 0.1, $\tau_0$ = 10, $\tau_s$ = 2) are plotted vs. the number of active ants in each hour. Red curve represents the line of best fit, with an effective mean excavation rate of 8.1 pellets/active ant/hour. \textbf{(B)} The number of pellets deposited over eight hours of simulation are plotted vs. the mean number of active ants across eight hours. Error bars represent standard deviation in number of active ants across all hours of the simulation. The red curve represents the line of best fit, with an effective mean excavation rate of 8.1 pellets/active ant/hour.}
    \label{fig:SI_sim_rates}
\end{figure}

Pellets transported per hour across all 880 hours of simulation data (for one set of parameters: $C_0$ = 10, $R$ = 0.1, $\tau_0$ = 10, $\tau_s$ = 2) are shown in Figure \ref{fig:SI_sim_rates}A. Least squares regression yields a resultant fit rate of $\approx$8.1  pellets per active ant per hour (Figure \ref{fig:SI_sim_rates}A). 
Figure \ref{fig:SI_sim_rates}B compares total pellets transported over eight hours to mean number of active ants over eight hours.

\subsection*{Summary of all experimental trials parsed via tracking algorithm}
\begin{figure}
    \centering
    \includegraphics[width=0.75\linewidth]{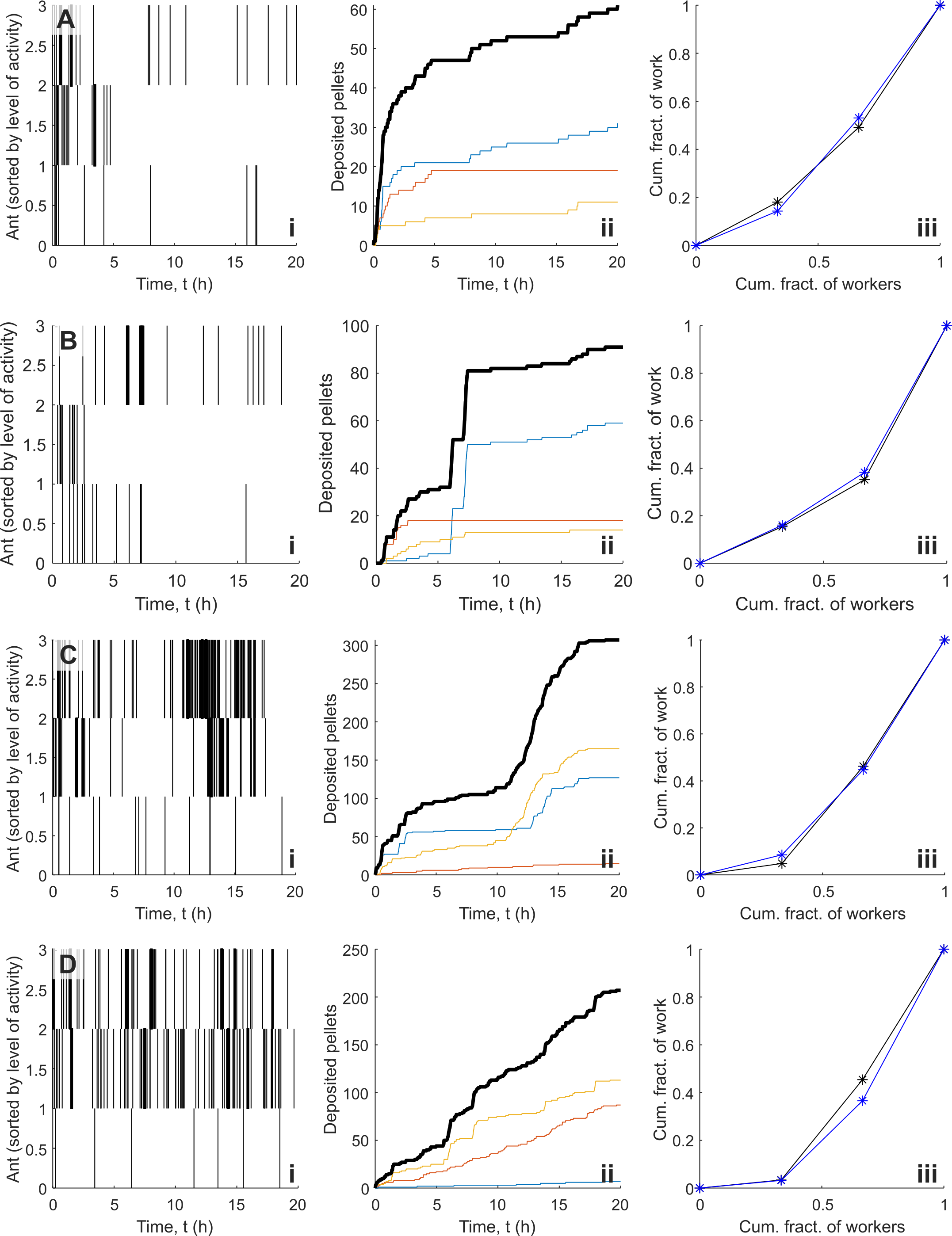}
    \caption{\textbf{Experimental results, 3 ant trials}. Results for four different experimental trials, each containing three ants. Each row of plots \textbf{(A-D)} represents a single experimental trial. Leftmost plots \textbf{(i)} represent ant activity over the trial duration, where a black dash indicates a grain transported at that time instance. Each row corresponds to a different ant, with the top row corresponding to the most active ant. Center plots \textbf{(ii)} show the number of deposited pellets, as estimated by the tracking algorithm, for each ant across 20 hours. Colored lines represent individual ant contributions, while the black curve represents the total pellets deposited. Rightmost plots \textbf{(iii)} are Lorenz curves, representing cumulative fraction of grains moved, relative to cumulative fraction of workers involved, for the first 8 hours (blue) and full 20 hour trial (black).}
    \label{fig:SI_3ant}
\end{figure}

\begin{figure}
    \centering
    \includegraphics[width=0.75\linewidth]{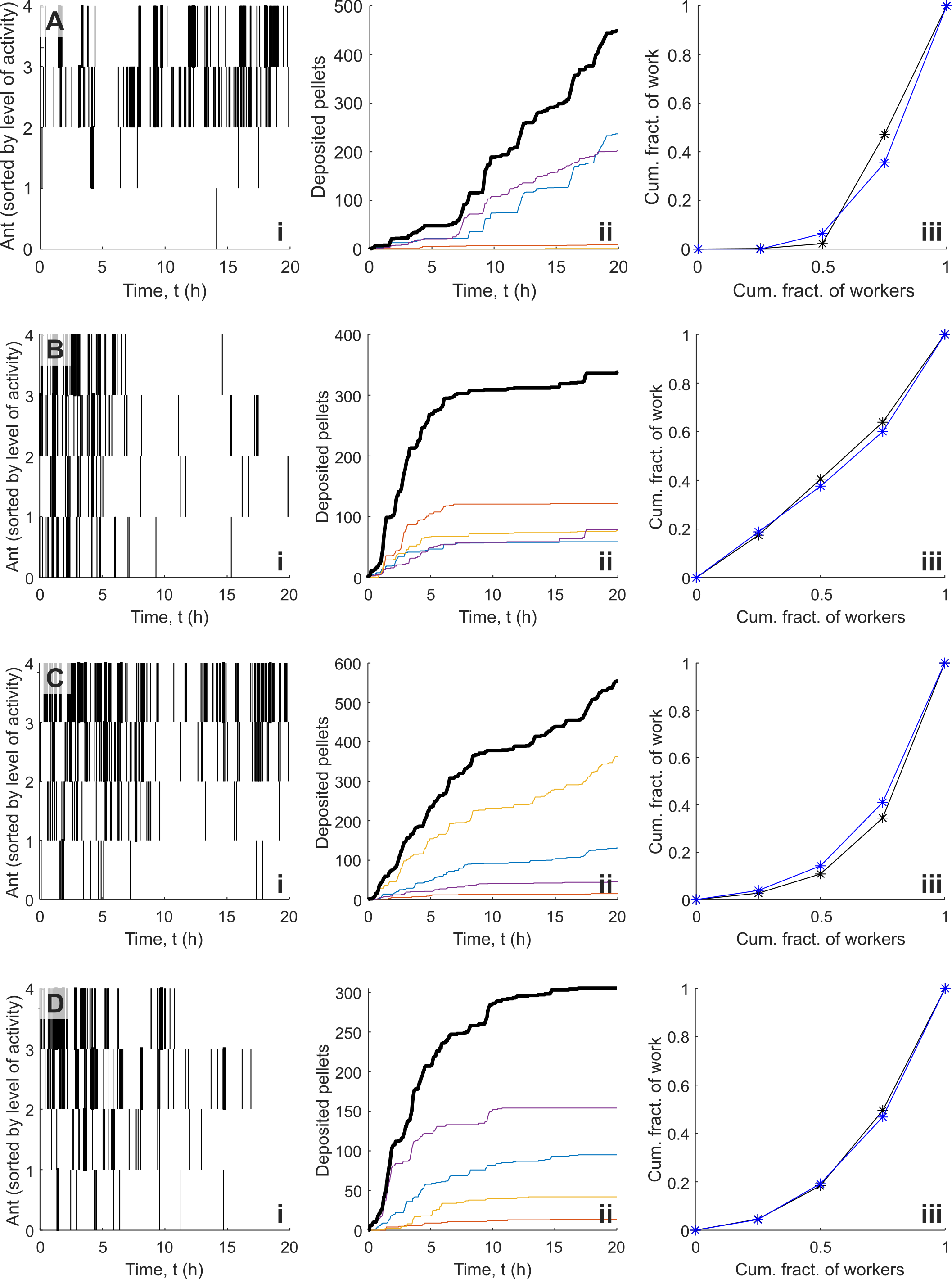}
    \caption{\textbf{Experimental results, 4 ant trials}. Results for four different experimental trials, each containing four ants. Each row of plots \textbf{(A-D)} represents a single experimental trial. Leftmost plots \textbf{(i)} represent ant activity over the trial duration, where a black dash indicates a grain transported at that time instance. Each row corresponds to a different ant, with the top row corresponding to the most active ant. Center plots \textbf{(ii)} show the number of deposited pellets, as estimated by the tracking algorithm, for each ant across 20 hours. Colored lines represent individual ant contributions, while the black curve represents the total pellets deposited. Rightmost plots \textbf{(iii)} are Lorenz curves, representing cumulative fraction of grains moved, relative to cumulative fraction of workers involved, for the first 8 hours (blue) and full 20 hour trial (black). }
    \label{fig:SI_4ant}
\end{figure}

\begin{figure}
    \centering
    \includegraphics[width=0.75\linewidth]{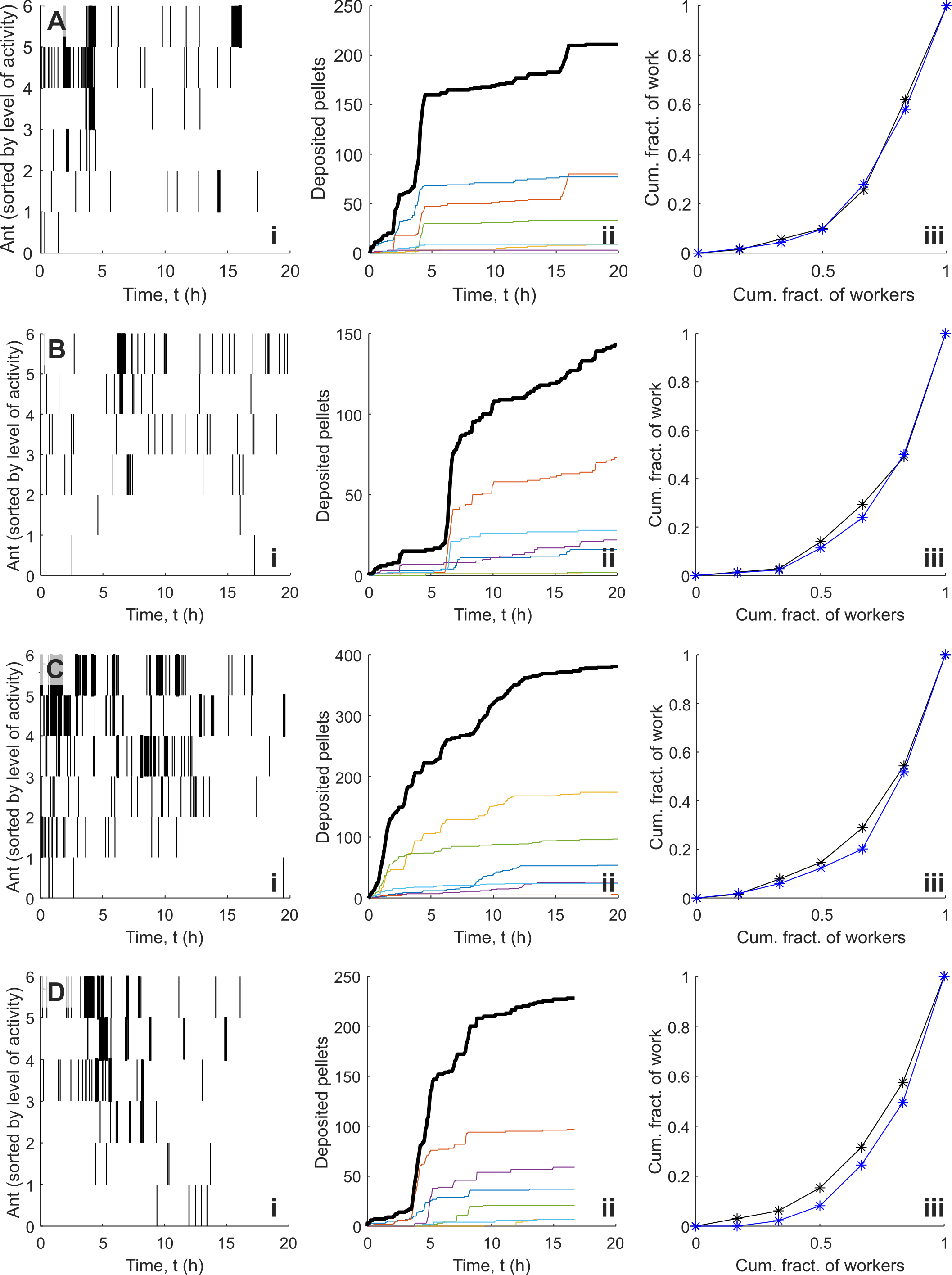}
    \caption{\textbf{Experimental results, 6 ant trials}. Results for four different experimental trials, each containing six ants. Each row of plots \textbf{(A-D)} represents a single experimental trial. Leftmost plots \textbf{(i)} represent ant activity over the trial duration, where a black dash indicates a grain transported at that time instance. Each row corresponds to a different ant, with the top row corresponding to the most active ant. Center plots \textbf{(ii)} show the number of deposited pellets, as estimated by the tracking algorithm, for each ant across 20 hours. Colored lines represent individual ant contributions, while the black curve represents the total pellets deposited. Rightmost plots \textbf{(iii)} are Lorenz curves, representing cumulative fraction of grains moved, relative to cumulative fraction of workers involved, for the first 8 hours (blue) and full 20 hour trial (black).}
    \label{fig:SI_6ant}
\end{figure}

\begin{figure}
    \centering
    \includegraphics[width=0.75\linewidth]{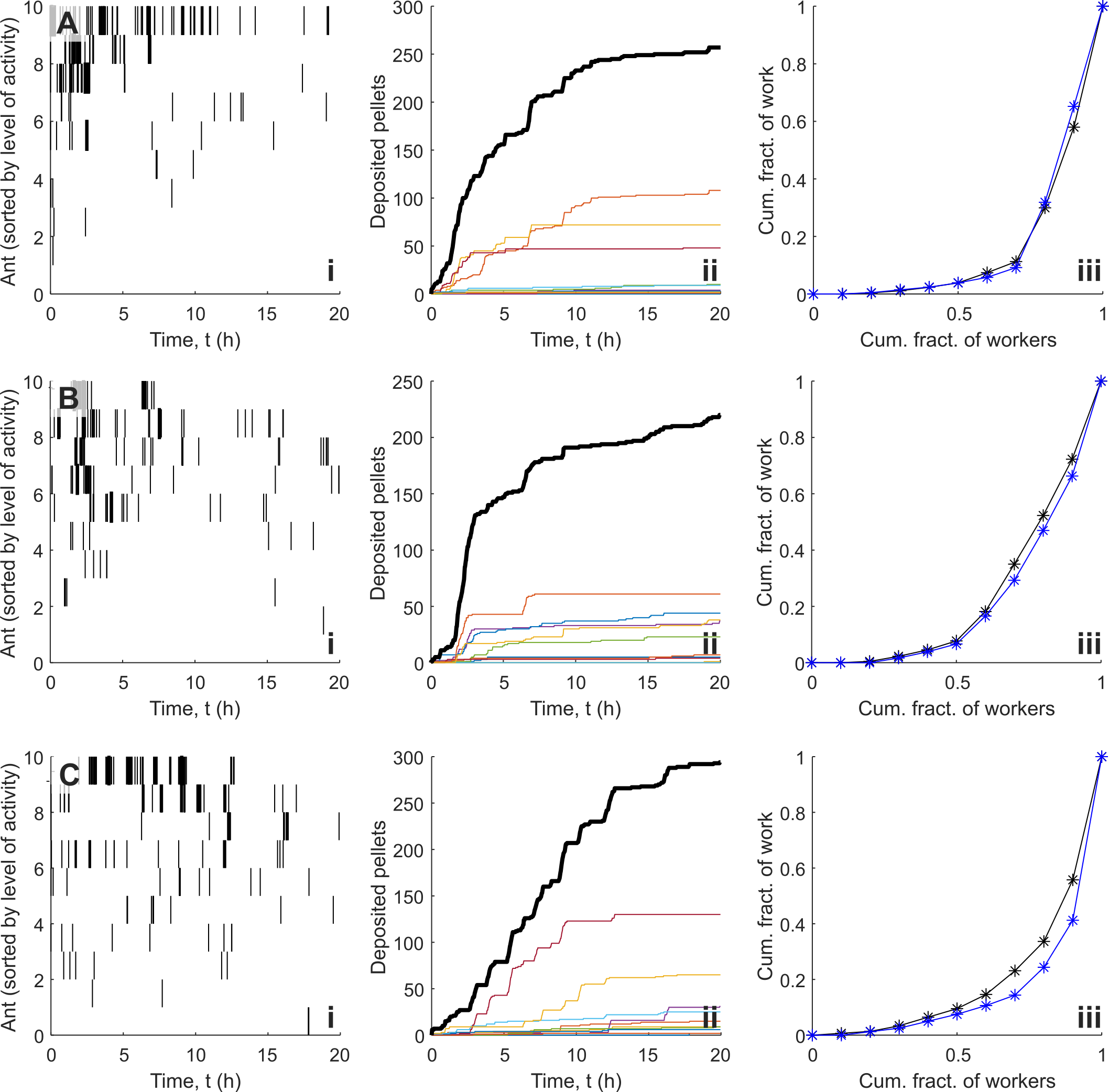}
    \caption{\textbf{Experimental results, 10 ant trials}. Results for three different experimental trials, each containing ten ants. Each row of plots \textbf{(A-C)} represents a single experimental trial. Leftmost plots \textbf{(i)} represent ant activity over the trial duration, where a black dash indicates a grain transported at that time instance. Each row corresponds to a different ant, with the top row corresponding to the most active ant. Center plots \textbf{(ii)} show the number of deposited pellets, as estimated by the tracking algorithm, for each ant across 20 hours. Colored lines represent individual ant contributions, while the black curve represents the total pellets deposited. Rightmost plots \textbf{(iii)} are Lorenz curves, representing cumulative fraction of grains moved, relative to cumulative fraction of workers involved, for the first 8 hours (blue) and full 20 hour trial (black). }
    \label{fig:SI_10ant}
\end{figure}

Here, we report the data collected for all experimental trials with 10 or fewer ants (Figures \ref{fig:SI_3ant},\ref{fig:SI_4ant},\ref{fig:SI_6ant},\ref{fig:SI_10ant}). Each row (A-D) represents a different trial and 20 hours of data are shown for each. All results shown here were generated via custom MATLAB tracking algorithms. 

\subsection*{Results of full parameter sweeps in CA simulation}
\begin{figure}
    \centering
    \includegraphics[width=0.85\linewidth]{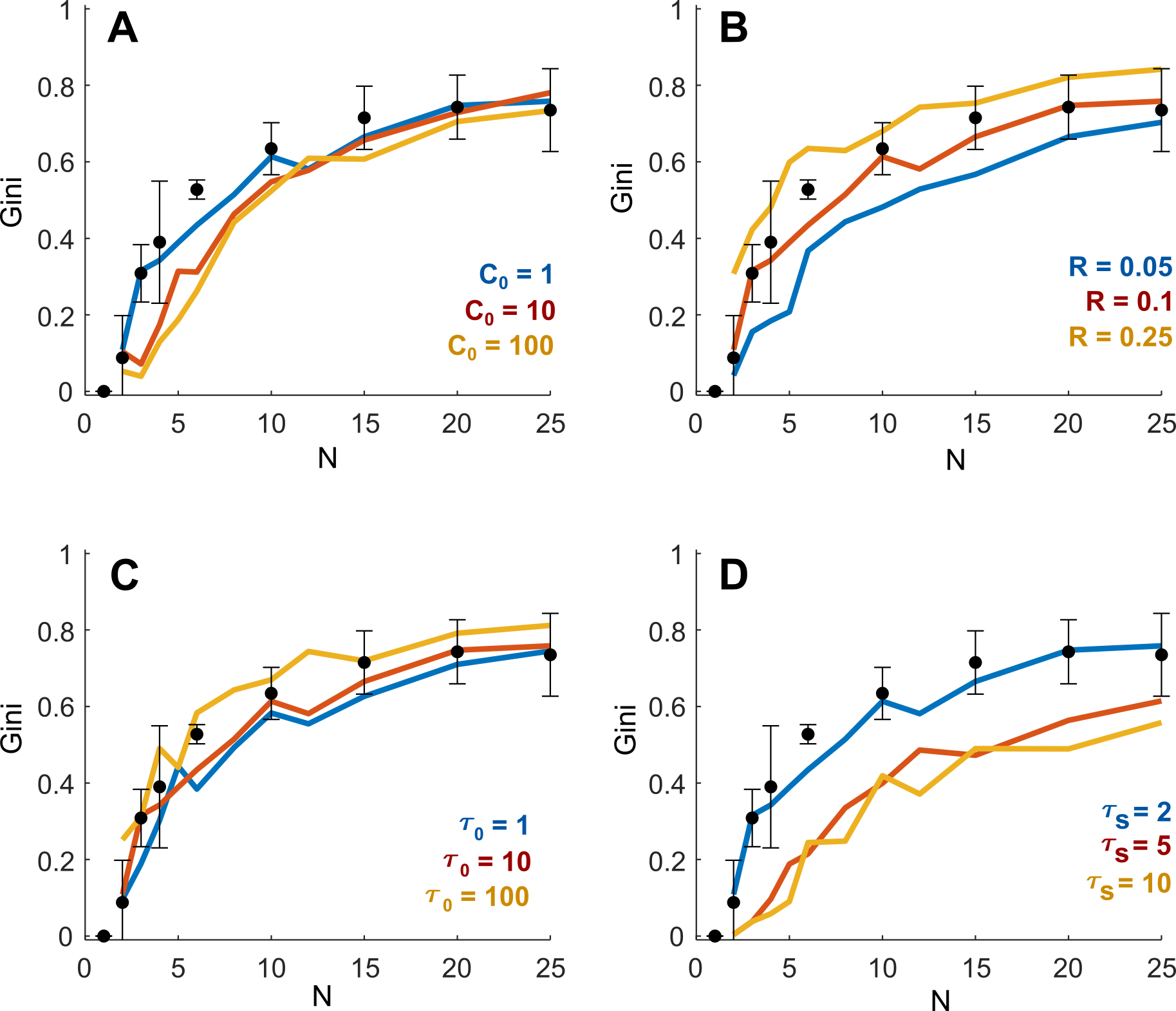}
    \caption{\textbf{Effect of all parameters on CA simulation Gini coefficient estimates}. Each plot represents the effect of a single parameter on Gini coefficient trends; each line represents the mean Gini coefficients over 10 trials at each group size, $N$. In all plots (A)-(D), the default parameters are: $C_0$ = 1, $R$ = 0.1, $\tau_0$ = 10, $\tau_s$ = 2. 
    \textbf{(A)} The effect of crowding sensitivity, $C_0$. \textbf{(B)} The effect of intrinsic work-rest ratio, $R$. 
    \textbf{(C)} The effect of timescale for the tolerance of a work-to-rest imbalance, $\tau_0$. 
    \textbf{(D)} The effect of memory (timescale over which success rate is calculated), $\tau_s$. 
    }
    \label{fig:SI_gini_all_params}
\end{figure}

\begin{figure}
    \centering
    \includegraphics[width=0.85\linewidth]{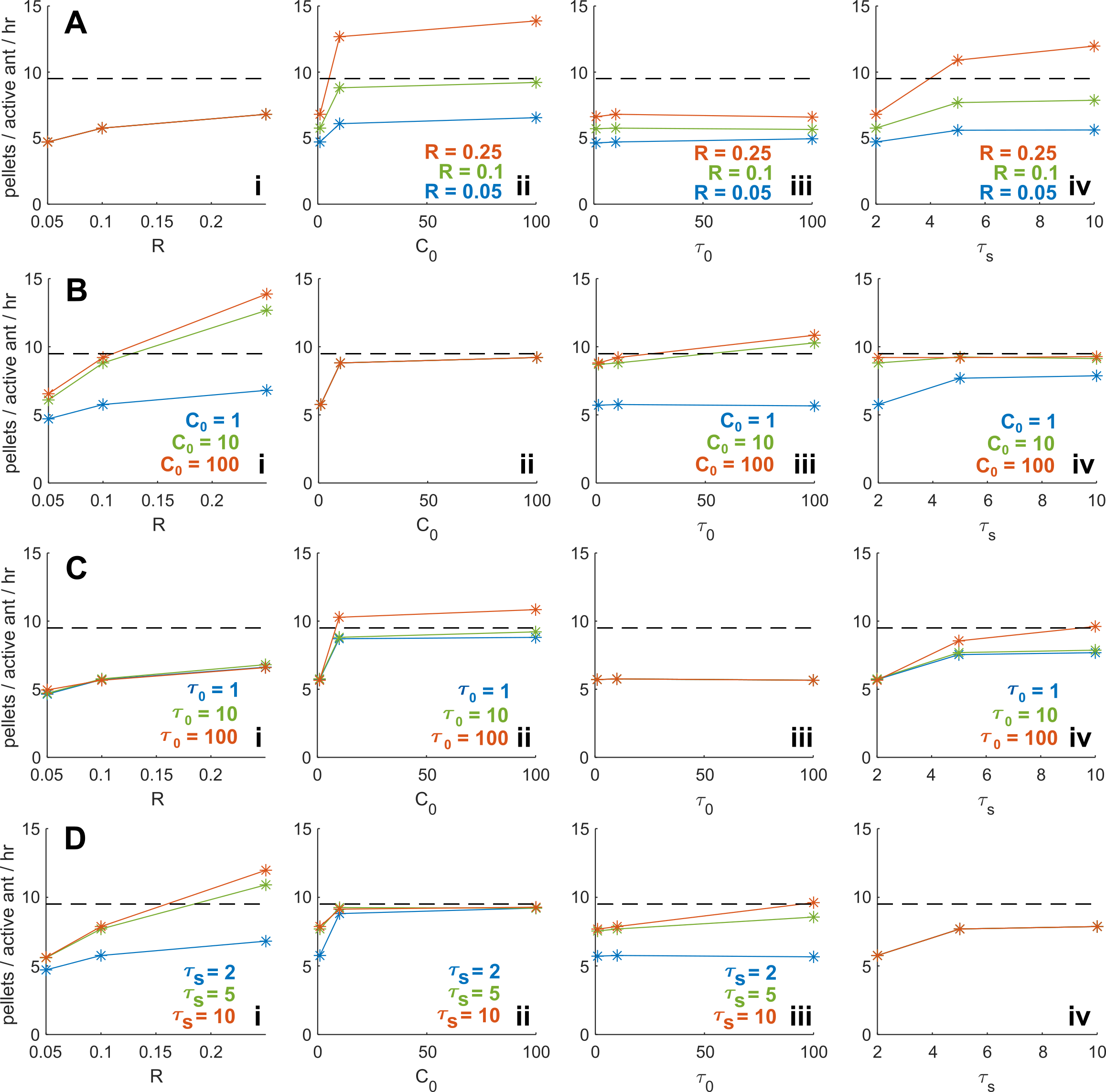}
    \caption{\textbf{Effect of all parameters on CA simulation excavation rates}. Each plot (A-D) represents the effect of a single parameter on resultant excavation rates, and each subplot (i - iv) shows how varying another parameter affects these trends. Three values of each parameter were tested, and asterisks represent the estimated excavation rate for each parameter combination (based on the method described above). In all plots (A)-(D), unless otherwise indicated, the default parameters are: $C_0$ = 1, $R$ = 0.1, $\tau_0$ = 10, $\tau_s$ = 2. 
    \textbf{(A)} The effect of intrinsic work-rest ratio, $R$.
    \textbf{(B)} The effect of crowding sensitivity, $C_0$.  
    \textbf{(C)} The effect of timescale for the tolerance of a work-to-rest imbalance, $\tau_0$. 
    \textbf{(D)} The effect of memory (timescale over which success rate is calculated), $\tau_s$. }
    \label{fig:SI_rates_allparams}
\end{figure}

Only some CA model parameters have a measurable effect on resultant Gini coefficient trends. Here, we vary each parameter individually and report the change in mean Gini coefficient for each combination of parameters. All results are shown in Figure \ref{fig:SI_gini_all_params}A-D. Trends in excavation rates for all combinations of simulation parameters are shown in Figure \ref{fig:SI_rates_allparams}.

\begin{table}[h]
\centering
\begin{tabular}{|l|c|c|c|}
\hline
\textbf{Model parameters} & \textbf{Value / Range} & \textbf{Units} & \textbf{Source} \\
\hline
Reversal probability & 0.34 &  & \cite{aguilar_collective_2018} \\
\hline
Tunnel width & 2 & cells & \cite{aguilar_collective_2018} \\
\hline
average time spent excavating & 8.5 & seconds & \cite{avinery_agitated_2023} \\
\hline
 & 2 & timesteps & \cite{avinery_agitated_2023} \\
\hline
number of ant trips per cm excavated & 206 &  & \cite{avinery_agitated_2023} \\
\hline
number of ant trips per cell excavated & 40 &  & \cite{avinery_agitated_2023} \\
\hline
Cell width in CA simulation & 1.1 & mm & \cite{gravish_effects_2012} \\
\hline
Cell length in CA simulation & 3.95 & mm & \cite{gravish_effects_2012} \\
\hline
Average ant walking speed & 0.27 & BL/sec & \cite{avinery_agitated_2023} \\
\hline
Time per simulation timestep & 3.70 & sec & \cite{avinery_agitated_2023} \\
\hline
Total simulation time & 8 & hrs &  \\
\hline
 & 7776 & timesteps &  \\
\hline
Lateral motion probability & 0.52 &  & \cite{aguilar_collective_2018} \\
\hline
time to drop pellet & 10 & seconds & \cite{aguilar_collective_2018} \\
\hline
 & 3 & timesteps & \cite{aguilar_collective_2018} \\
\hline
R (default work/rest imbalance) & 0.05 -- 0.25 &  & \cite{avinery_agitated_2023} \\
\hline
$C_0$ (crowding sensitivity parameter) & 1 -- 100 &  & new \\
\hline
$\tau_0$ (timescale for the tolerance of a work-to-rest imbalance) & 1 -- 100 & timesteps & \cite{avinery_agitated_2023} \\
\hline
$\tau_s$ (period over which success rate is calculated) & 2 -- 10 & timesteps & new \\
\hline
\end{tabular}
\caption{\textbf{Simulation Parameters}. All parameters used for implementation of the cellular automata are presented. We indicate the exact quantity if a single parameter value was used, or a range if multiple parameters were tested. Many parameter values were established in prior work, by either Aguilar et al. \cite{aguilar_collective_2018},  Avinery et al. \cite{avinery_agitated_2023}, or Gravish et al. \cite{gravish_effects_2012}. Several new parameters are introduced in this CA model, and their values are indicated.  }
\label{fig:sim_table}
\end{table}

\newpage

\end{document}